\documentclass[epj]{svjour}
\usepackage{graphics}
\usepackage{epsfig}
\begin{document}
%
\title{On particle production for high energy neutrino beams}
\author{M.~Bonesini\inst{1} \and
        A.~Marchionni\inst{2} 
           \thanks{On leave of absence from INFN, Sezione di Firenze,
                   Largo E. Fermi 2, I-50125, Florence, Italy} \and
	F.~Pietropaolo\inst{3} 
           \thanks{On leave of absence from INFN, Sezione di Padova, 
                   Via Marzolo 8, I-35131, Padua, Italy} \and
        T.~Tabarelli de Fatis\inst{1}
}  

\institute{INFN, Sezione di Milano,  Via Celoria 16, I-20133 Milan, Italy
    \and   Fermi National Accelerator Laboratory,  Batavia, Illinois 60510
    \and   CERN, CH-1211 Geneva 23, Switzerland}
%
\date{\today}
%
\abstract{
Analytical formulae for the calculation of secondary particle  
yields in p-A interactions are given. These formulae can be of great
practical importance for fast calculations of neutrino fluxes and for
designing new neutrino beam-lines. The formulae  are based on a
parameterization of the inclusive invariant cross sections for
secondary particle production measured in p-Be interactions. Data
collected in different energy ranges and kinematic regions are
used. The accuracy of the fit to the data with the empirical formulae 
adopted is within the experimental uncertainties.  Prescriptions to
extrapolate this parameterization to finite targets and to targets of
different materials are given.  The results obtained are then used as
an input for the simulation  of neutrino beams. We show that
our approach describes well the main characteristics of measured
neutrino spectra at CERN. Thus it may be used in fast simulations
aiming at the optimisation of the long-baseline neutrino beams at CERN 
and FNAL. In particular we will show our predictions for the CNGS beam 
from CERN to Gran Sasso. 
\PACS{
      {13.85.Ni}{Inclusive production with identified hadrons} \and
      {29.25.-t}{Particle sources and targets} \and
      {29.27.Eg}{Beam transport}
     } 
} 
\maketitle

\section{Introduction}
\label{intro}

A renewed interest in the predictions of absolute flux,  energy
distribution and flavour composition of neutrino beams produced at
accelerators is motivated by the long-baseline neutrino beams in
construction at CERN (CNGS \cite{CNGS}) and FNAL (NuMI \cite{NUMI}) 
and derived from the decays in flight of mesons generated by 400 GeV/c 
and 120 GeV/c proton interactions on graphite, respectively.

The available experimental data on particle yields from high energy
proton interactions on light nuclei targets, such as the ones used
in the production of neutrino beams, are not extensive and one is
often faced with the problem of extrapolating to different  target
materials and shapes and to different incident proton energies.

Motivated by the need to estimate the neutrino flux at the West Area
Neutrino Facility at CERN \cite{WANF}, a consistent set of
measurements of particle yields on Beryllium targets of different
lengths were performed at CERN by the NA20 Coll. \cite{NA20} and
subsequently by the NA56/SPY Coll. \cite{SPY1,SPY2}, with 400 GeV/c 
and 450 GeV/c incident protons, respectively. Secondary particles 
($\pi^+$, $\pi^-$, $K^+$, $K^-$, $p$, $\overline{p}$) were measured 
in a transverse momentum range from 0 to 600 MeV/c, with the NA20
measurements in the high momentum range (60 GeV/c $< p <$ 300 GeV/c) 
and with NA56/ SPY extending to the low momentum region 
(7 GeV/c $< p <$ 135 GeV/c), thus covering also the relevant 
kinematic range of secondary particles for the planned long-baseline 
(LBL) neutrino beams.

In this work, in order to make these measurements of general
applicability, the measured single-particle yields on Beryllium 
have been converted to single-particle invariant cross sections, 
correcting for target efficiency and tertiary particle production.  
The resulting data has been empirically parameterized as a function of
the transverse momentum ($p_T$) and the scaling variable
$x_R=E^*/E^*_{max}$, defined as the ratio of the energy of the
detected particle in the centre-of-momentum frame and the maximum 
energy kinematically available to the detected particle, with a
formula based on general physical arguments. The choice of the
variables used in the description of the invariant cross section is 
motivated by the approximate scaling behaviour of the invariant cross 
section, when expressed in terms of these variables.
A satisfactory comparison of this empirical parameterization, based on 
data collected with 400 GeV/c and 450 GeV/c protons, with invariant 
cross section measurements with 100 GeV/c protons shows that such 
approach might be useful also for the prediction of the NuMI neutrino 
beam at FNAL.
 
Extension of the parameterization to target materials other than
Beryllium is possible by the known dependence of the invariant cross
section on the  atomic number as shown in the measurements of Barton 
et al. \cite{Bar87}. This is particularly relevant for the planned
long-baseline neutrino experiments at CERN and FNAL, both planning to
use a graphite target.

Such parameterization of the invariant cross section for proton-nucleus
interactions could be implemented in a full Monte Carlo simulation of
the target geometry. On the other hand, we show that it is possible to
compute  analytically particle yields from a given target by taking
into account the geometry of the target and the contribution of
cascade processes.  Based on the NA56/SPY measurements of particle yields
from targets of different  lengths, an empirical formula describing
tertiary particle production is derived.

This approach, coupled to the tracking of the produced secondary
hadrons  through the neutrino beam-line elements (magnetic lenses,
collimators and decay tunnel), has been implemented in a fast Monte
Carlo for the explicit calculation of neutrino beam fluxes. Effects
due to reinteractions in the beam-line material have  been
parameterized with good approximation. A comparison with the
neutrino and anti-neutrino spectra measured by the CHARM II 
experiment at CERN is shown. Predictions for the planned CNGS 
long-baseline neutrino beam are given.

With respect to the standard approach using complex Monte Carlo
simulations based on hadronic cascade codes, such as GEISHA
\cite{GHEI} or FLUKA92 \cite{FL92}, as implemented in the GEANT
package \cite{GEAN}, CALOR95 \cite{CA95}, the stand-alone version of
FLUKA \cite{FONE} or MARS \cite{MARS}, a simple empirical
parameterization of particle yields as described here has the advantage
of a more transparent physical input to the calculations. Moreover,
when coupled to the  tracking of the secondary hadrons from the
target, it becomes a fast neutrino  Monte Carlo generator with the
obvious advantage of enhanced statistics.  Detailed studies of
systematic effects due for example to horn misalignments in the
neutrino beam-line are made possible even with limited computer time.

\section{Parameterization of p-Be data}

An empirical formula for particle production, based on the yields from
thick Be targets measured by NA20, has long ago been proposed by
Malensek \cite{Mal81}.  This has already offered an alternative
approach to neutrino beam computations \cite{NTeV}. Extensions of this
parameterization to include effects related to the target geometry have
been proposed \cite{Bon96}. Malensek's parameterization, however, fails
to describe the particle production in the low momentum region covered 
by NA56/SPY measurements and cannot be stretched to consistently fit the 
whole set of available data on p-Be interactions. 
An alternative parameterization of NA20 data on particle yields,
proposed by NA20 \cite{NA20}, gives an even less accurate description
of the NA56/ SPY data.

At variance with these works, the particle yields measured by the NA20
and NA56/SPY collaborations have been converted to inclusive invariant
cross sections for particle production, which have then been
parameterized. 

\subsection{The data}
\label{ics}

In Ref. \cite{SPY2}, inclusive invariant cross sections for particle
production in the forward direction in p-Be interactions have been
reported by the NA56/SPY Collaboration. We have used the same method
discussed in that paper to derive inclusive invariant cross section at
all energies and angles from the published NA20 \cite{NA20} and NA56/SPY
\cite{SPY1,SPY2} data on particle yields in p-Be interactions.

The inclusive invariant particle production cross section is related
to the measured yields from targets of finite length by:
\begin{equation}
\label{I2Y}
E \times \frac{d^3 \sigma}{d p^3} = (100 \cdot Y) \frac{E}{p^3}
\frac{A}{N_0 \rho \lambda_p F(L)}
\end{equation}
where $Y$ is the yield per incident proton per sr~($\Delta p/p$ \%);
$A$, $\rho$, $L$ the atomic mass number, density and length of the
target; $N_0$ the Avogadro's number; $E$ and $p$ the energy and
momentum of the secondary particle, respectively.  The target
efficiency $F(L)$ can be estimated assuming that the produced
secondary particles if absorbed in the target do not generate
additional particles.  Under this approximation and for forward
production, one has
\begin{equation}
\label{targeff}
F(L) = \int_0^L \frac{dz}{\lambda_p} e^{-z/\lambda_p}
e^{-(L-z)/\lambda_s}
\end{equation}
representing the convolution of the probability that the primary
proton interacts between $z$ and $z+dz$ in the target and that the
produced secondary particle escapes from the target ($\lambda_p$ and
$\lambda_s$ are the effective mean free paths of primary 
and secondary particles respectively).  This model neglects the 
production of tertiaries.  This effect can be accounted for, in a 
model independent way, by estimating the invariant cross sections 
for various target lengths and then extrapolating the results to 
zero target thickness \cite{SPY2}. 

This procedure also allows to estimate at each momentum the
contribution to the total yield from finite targets due to tertiary
particles.

For particle production at angles different from zero, the method is
easily generalised by taking into account the target geometry in
eq. (\ref{targeff}).  At angles different from zero, however, the NA56/SPY
collaboration has measured the production yields only at a fixed
target length.  In order to derive inclusive invariant cross sections
from those data points, we have made the additional assumption that
the correction for tertiary production be angle independent. This has
been verified to be approximately true in the kinematic region covered 
by NA20 data, collected with a planar target setup identical to the
one adopted in the NA56/SPY experiment.

In this way we have derived the single-particle inclusive invariant 
cross sections at all available energies and production angles from 
NA20 and NA56/SPY data. Table \ref{data} summarizes our results. Errors
include the experimental errors and the uncertainty on the 
extrapolation procedure.

\begin{table*}[htb]
\begin{center} 
\begin{tabular}{cl|cccccc|c}
\hline\noalign{\smallskip}
           &\multicolumn{1}{c|}{$p_T$}  & \multicolumn{6}{c|}{$(E\times {d^3\sigma}/{dp^3})_{pBe\to hX}$ (mb/GeV$^2/c^3$)} &   \\ 
 $x_{Lab}$ &\multicolumn{1}{c|}{(GeV/c)}& $\pi^+$ &  $\pi^-$  &      $K^+$         &        $K^-$       &       $p$          &   $\overline{p}$  &    Ref.    \\  
\noalign{\smallskip}\hline\noalign{\smallskip}
0.015&       0.  &    532.6$\pm$55.7   &    490.0$\pm$55.6    &    29.9$\pm$3.4    &    26.6$\pm$4.7    &    21.0$\pm$3.0    &     6.1$\pm$0.9   &\cite{SPY2} \\
\noalign{\smallskip}\hline\noalign{\smallskip}
0.022&     0.     &    433.4$\pm$38.3   &          --         &    30.2$\pm$3.1    &          --        &    16.8$\pm$1.6    &         --        &\cite{SPY2} \\
\noalign{\smallskip}\hline\noalign{\smallskip}
0.033&     0.     &    347.9$\pm$35.2   &    289.2$\pm$30.3   &    27.1$\pm$3.0    &    21.4$\pm$2.2    &    16.7$\pm$1.9    &    8.5$\pm$0.9    &\cite{SPY2} \\
0.033&     0.075  &    369.9$\pm$37.4   &    293.6$\pm$30.8   &    27.5$\pm$2.7    &    20.0$\pm$2.0    &    16.3$\pm$1.9    &    8.4$\pm$0.9    &\cite{SPY2} \\ 
0.033&     0.150  &    334.1$\pm$33.8   &          --         &    25.1$\pm$2.7    &          --        &    15.8$\pm$1.8    &         --        &\cite{SPY2} \\
0.033&     0.225  &    242.2$\pm$24.5   &    200.1$\pm$21.0   &    19.8$\pm$2.1    &    15.4$\pm$1.5    &    13.9$\pm$1.6    &    6.8$\pm$0.7    &\cite{SPY2} \\
0.033&     0.3375 &    127.7$\pm$12.9   &          --         &    14.2$\pm$1.6    &          --        &    10.5$\pm$1.2    &         --        &\cite{SPY2} \\
0.033&     0.450  &     68.4$\pm$6.9    &          --         &     8.8$\pm$0.9    &          --        &     7.8$\pm$1.0    &         --        &\cite{SPY2} \\
\noalign{\smallskip}\hline\noalign{\smallskip}
0.045&     0.     &    293.0$\pm$24.8   &          --         &    27.0$\pm$2.5    &          --        &    18.0$\pm$1.6    &         --        &\cite{SPY2} \\
\noalign{\smallskip}\hline\noalign{\smallskip}
0.067&     0.     &    222.0$\pm$14.9   &          --         &    22.2$\pm$1.6    &          --        &    18.9$\pm$1.5    &         --        &\cite{SPY2} \\
\noalign{\smallskip}\hline\noalign{\smallskip}
0.090&     0.     &    169.7$\pm$9.7    &    137.8$\pm$7.8    &    18.4$\pm$1.1    &    11.5$\pm$0.8    &    19.2$\pm$1.2    &    5.2$\pm$0.3    &\cite{SPY2} \\
0.090&     0.075  &    176.1$\pm$9.9    &    139.3$\pm$7.8    &    18.0$\pm$1.2    &    11.5$\pm$0.8    &    19.0$\pm$1.2    &    5.3$\pm$0.3    &\cite{SPY2} \\
0.090&     0.150  &    186.4$\pm$10.5   &    135.3$\pm$7.6    &    17.0$\pm$1.1    &    10.5$\pm$0.7    &    18.0$\pm$1.1    &    4.9$\pm$0.3    &\cite{SPY2} \\
0.090&     0.225  &    160.2$\pm$9.1    &    114.5$\pm$6.4    &    14.7$\pm$1.0    &     9.6$\pm$0.7    &    16.6$\pm$1.0    &    4.8$\pm$0.3    &\cite{SPY2} \\
0.090&     0.450  &     58.2$\pm$3.3    &     44.0$\pm$2.5    &     7.7$\pm$0.5    &     4.6$\pm$0.3    &    10.0$\pm$0.6    &    2.7$\pm$0.2    &\cite{SPY2} \\
0.090&     0.600  &     22.8$\pm$1.3    &     20.3$\pm$1.1    &     3.9$\pm$0.2    &     2.6$\pm$0.2    &     6.0$\pm$0.4    &    1.6$\pm$0.1    &\cite{SPY2} \\
\noalign{\smallskip}\hline\noalign{\smallskip}
0.150&     0.     &    111.1$\pm$7.0    &     64.5$\pm$4.7    &    11.7$\pm$0.9    &     5.5$\pm$0.4    &    24.4$\pm$1.8    &    2.4$\pm$0.2    &\cite{SPY2} \\
0.150&     0.     &    105.0$\pm$6.5    &     70.5$\pm$4.5    &    10.7$\pm$0.7    &    5.72$\pm$0.46   &    22.3$\pm$1.5    &   2.77$\pm$0.21   &\cite{NA20} \\
0.150&     0.500  &     33.8$\pm$2.0    &     22.3$\pm$1.4    &     4.7$\pm$0.3    &    2.26$\pm$0.12   &    10.5$\pm$0.7    &   1.23$\pm$0.07   &\cite{NA20} \\
\noalign{\smallskip}\hline\noalign{\smallskip}
0.300&     0.     &     64.0$\pm$4.6    &          --         &     5.0$\pm$0.4    &          --        &    40.6$\pm$3.0    &         --        &\cite{SPY2} \\
0.300&     0.     &     68.1$\pm$4.7    &    23.2$\pm$1.4     &    5.30$\pm$0.37   &    1.51$\pm$0.09   &    43.1$\pm$3.0    &  0.493$\pm$0.030  &\cite{NA20} \\
0.300&     0.300  &     35.4$\pm$2.4    &    14.1$\pm$0.9     &    4.13$\pm$0.28   &    1.06$\pm$0.07   &    32.0$\pm$2.1    &  0.343$\pm$0.025  &\cite{NA20} \\    
0.300&     0.500  &     17.1$\pm$1.2    &     8.4$\pm$0.5     &    2.60$\pm$0.17   &    0.66$\pm$0.04   &    20.4$\pm$1.7    &  0.254$\pm$0.016   &\cite{NA20} \\
\noalign{\smallskip}\hline\noalign{\smallskip}
0.500&     0.     &     14.6$\pm$4.3    &    5.00$\pm$0.29    &    1.70$\pm$0.50   &   0.226$\pm$0.015  &    55.6$\pm$13.8   &(3.56$\pm$0.21)$\times$10$^{-2}$&\cite{NA20} \\     
0.500&     0.500  &     4.49$\pm$0.30   &    1.64$\pm$0.09    &    0.85$\pm$0.06   &   0.088$\pm$0.006  &    20.4$\pm$1.7    &(1.39$\pm$0.08)$\times$10$^{-2}$&\cite{NA20} \\
\noalign{\smallskip}\hline\noalign{\smallskip}
0.750&     0.     &           --        &    0.27$\pm$0.02    &          --      &(2.82$\pm$0.17)$\times$10$^{-3}$&     --      &(1.32$\pm$0.07)$\times$10$^{-4}$&\cite{NA20} \\
0.750&     0.500  &     0.50$\pm$0.03   &   0.114$\pm$0.008   &   0.152$\pm$0.010&(1.36$\pm$0.11)$\times$10$^{-3}$&24.0$\pm$1.5 &(5.57$\pm$0.52)$\times$10$^{-5}$&\cite{NA20} \\
\noalign{\smallskip}\hline
\end{tabular} 
\caption{Inclusive invariant cross-section for $pBe \to hX$ production as a  
	 function of $x_{Lab}=p/p_{inc}$ and $p_T$, where $p$ and $p_{inc}$ 
	 are the momenta of the detected particle and of the incident
         proton in the laboratory reference frame. 
         Data on single-particle inclusive production yields of 
         Ref.~\cite{SPY2} and Ref.~\cite{NA20}, with incident proton 
         beams of 450 GeV/c and 400 GeV/c respectively, have been used.}
\label{data}
\end{center} 
\end{table*}

\subsection{The parameterization of inclusive invariant cross sections} 
\label{formula}

One of the goals of this analysis is to adopt a simple functional form
for inclusive particle production which will be appropriate for
extrapolation to different centre of mass energies and/or secondary
particle momenta.

Feynman has speculated in very general terms about the shape and energy 
dependence of inclusive processes \cite{Fey69}, suggesting an
approximate scaling behaviour of the single-particle inclusive invariant 
cross section when expressed in terms of the transverse momentum ($p_T$) 
and the longitudinal variable $x_F = 2 p^*_L / \sqrt{s}$, where $p^*_L$ 
and $\sqrt{s}$ are the longitudinal momentum of the detected particle and 
the total energy in the centre-of-momentum frame. A factorization in 
$x_F$ and $p_T$ of the invariant cross section has also been advocated 
as an experimental fact. 

An alternative scaling variable $x_R=E^*/E^*_{max}$, defined as the
ratio of the energy of the detected particle in the centre-of-momentum 
frame and the maximum energy kinematically available to the detected
particle, was suggested by Yen \cite{Yen74} and Taylor et
al. \cite{Tay76} and shown to greatly extend the range of validity 
of scaling at sub-asymptotic energies. As discussed in section 4.2, 
in the comparison of NA56/SPY and NA20 data to $pBe$ data collected at 
lower energies, we have not observed an improved scale invariance 
when $x_R$ is used. Nonetheless, we have adopted this variable because 
of two practical advantages, which translates in a simpler analytical 
parameterization: $x_R$ is always positive by construction and can 
never be zero, except for infinite energy in the centre-of-momentum 
frame. 

\begin{table*}[t]
\begin{center} 
\begin{tabular}{c|cccccccccc} 
\hline\noalign{\smallskip}
       &      $A$     &   $B$ &$\alpha$&$\beta$&      $a$     &      $b$     & $\gamma$ & $\delta$ & $r_0$ & $r_1$ \\  
       & (mb/GeV$^2$) &       &        &       & (GeV$^{-1}$) & (GeV$^{-2}$) &          &          &       &       \\
\noalign{\smallskip}\hline\noalign{\smallskip}
 $\pi$ & 62.3         & 1.57  & 3.45   & 0.517 & 6.10         &  --          & 0.153    & 0.478    & 1.05  & 2.65  \\
 $ K$  & 7.74         & --    & 2.45   & 0.444 & 5.04         &  --          & 0.121    & $2\gamma$& 1.15  & -3.17 \\
 $p$   & 8.69         & 12.3  &  --    &  --   & 5.77         & 1.47         &  --      &  --      &  --   &  --   \\
 $\overline{p}$& 5.20 & --    & 7.56   & 0.362 & 5.77         &  --          &  --      &  --      &  --   &  --   \\  
\noalign{\smallskip}\hline
\end{tabular} 
\caption{Values of the parameters corresponding to the best-fit of our
empirical parameterization of $\pi^{\pm}$ and $K^{\pm}$ inclusive production 
in p-Be interactions. Best-fit results on proton and anti-proton 
production data are also given (see text for details).} 
\label{tab:tabone}  
\end{center} 
\end{table*}

Early phenomenological analyses of ~$pp$~ data, based on the constituent
quark model, helped clarify that in hadron\-ic processes the produced
particles reflect the motion of the constituents \cite{Och77}. In 
particular, in the fragmentation region of the projectile, the
longitudinal momentum distribution of produced hadrons reflects the
momentum distribution of valence quarks inside the incident
hadron. This suggests an $x$ dependence of hadron production at 
large $x$ \footnote{At large $x_F$, $x_F$ and $x_R$ are equivalent.}
with a functional form similar to the one describing parton 
distributions. 
The functional shape of non-direct formation processes at small $x$, 
in which sea quarks are involved, however, is not easily described in 
this framework.

After some trials and considerations of the physical process, 
the following empirical parameterization of the inclusive invariant 
cross sections for positive sign secondary meson ($\pi^+$, $K^{+}$) 
production in p-Be interactions has been found to give a suitable 
description of data: 
\begin{eqnarray}
\label{eq1}
\nonumber
(E \times \frac{d^3 \sigma}{d p^3}) & = &
A (1-x_R)^{\alpha} (1+Bx_R) x_R^{-\beta} \times \\
  &  &  (1 + a'(x_R) p_T + b'(x_R) p_{T}^{2}) e^{-a'(x_R) p_T}
\end{eqnarray}
where $a'(x_R)=a/x_R^\gamma$ and $b'(x_R)=a^2/2 x_R^{\delta}$.

These formula assumes an approximate factorized scaling form in $x$
and $p_T$. The $(1-x)^{\alpha}$ behaviour at large $x$ is
theoretically motivated on the basis of quark counting rules 
\cite{Bla74,Gun79}. 
The $x^{-\beta}$ behaviour empirically accounts for the non-direct 
hadron formation mechanism at small $x$. 

The $p_T$ behaviour is modelled with the known exponential fall of soft
interactions and a polynomial behaviour to interpolate the low $p_T$
part of the spectrum.  The $x$ dependence of $a'(x)$ and $b'(x)$ is
introduced to parameterize the violation of $p_T$ invariance observed
in the data.  Models based on the parton structure of the hadrons
predict a $p_T^{-n}$ dependence of the cross section at large $p_T$,
where hard scattering processes take over.  A reasonable
parameterization of p-Be data with this functional form has not been
found. The possibility that the proposed parametric form fail to
describe particle production in the $p_T$ region not covered by
present data is acknowledged. 

The ratio {\it r} of positive to negative data ($\pi^+/\pi^-$ or
$K^+/K^-$) has been instead parameterized with the empirical formulae:
\begin{eqnarray}
\label{eq22} r(\pi) = r_0 \cdot (1+x_R)^{r_1} \\
\label{eq23}   r(K) = r_0 \cdot (1-x_R)^{r_1}
\end{eqnarray}
The shape of these ratios is supported by the phenomenological
analysis of $pp$ data of Ref. \cite{Och77}, showing that ${\it r}(\pi) 
\simeq 1$ for $x \simeq 0$ and rises to about 5 for $x \to 1$,
closely following the $u/d$ ratio of valence quarks in the projectile
proton, while ${\it r}(K)$ has a $(1-x)^{-3}$ behaviour for $x \to
1$. NA56/SPY and NA20 data only cover the fragmentation region of the
proton at large $x$ and the central region. At large $x$ a functional 
behaviour similar to the one exhibited by $pp$ data is expected. 

In order to keep the number of free parameters limited, positive and 
negative mesons are assumed to have the same $p_T$ distributions.
This has long been known to be only approximate in $pp$ data \cite{Allab}.

Table~\ref{tab:tabone} summarizes the results of our best-fit to the
data. As indicated in the table, some of the parameters have been 
fixed in the fitting procedure, since they appeared to be
redundant. In the $K^{\pm}$ fit, $\delta = 2 \gamma$ was chosen 
and $B$ was set to zero, since its fitted value was found to be 
consistent with zero within errors. 

\begin{figure*}
  \begin{center}
  \mbox{\epsfig{file=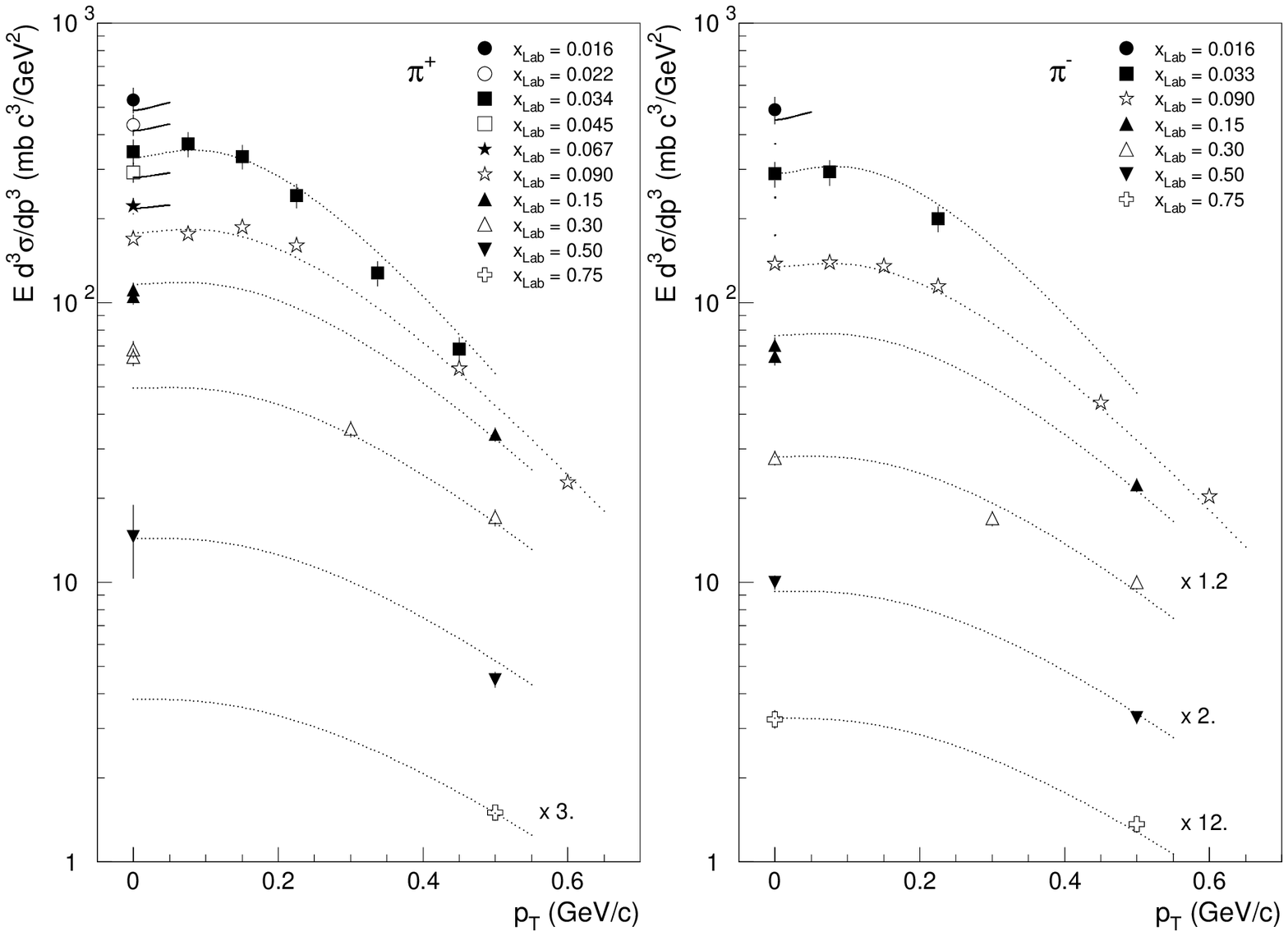,width=0.8\linewidth}}
  \mbox{\epsfig{file=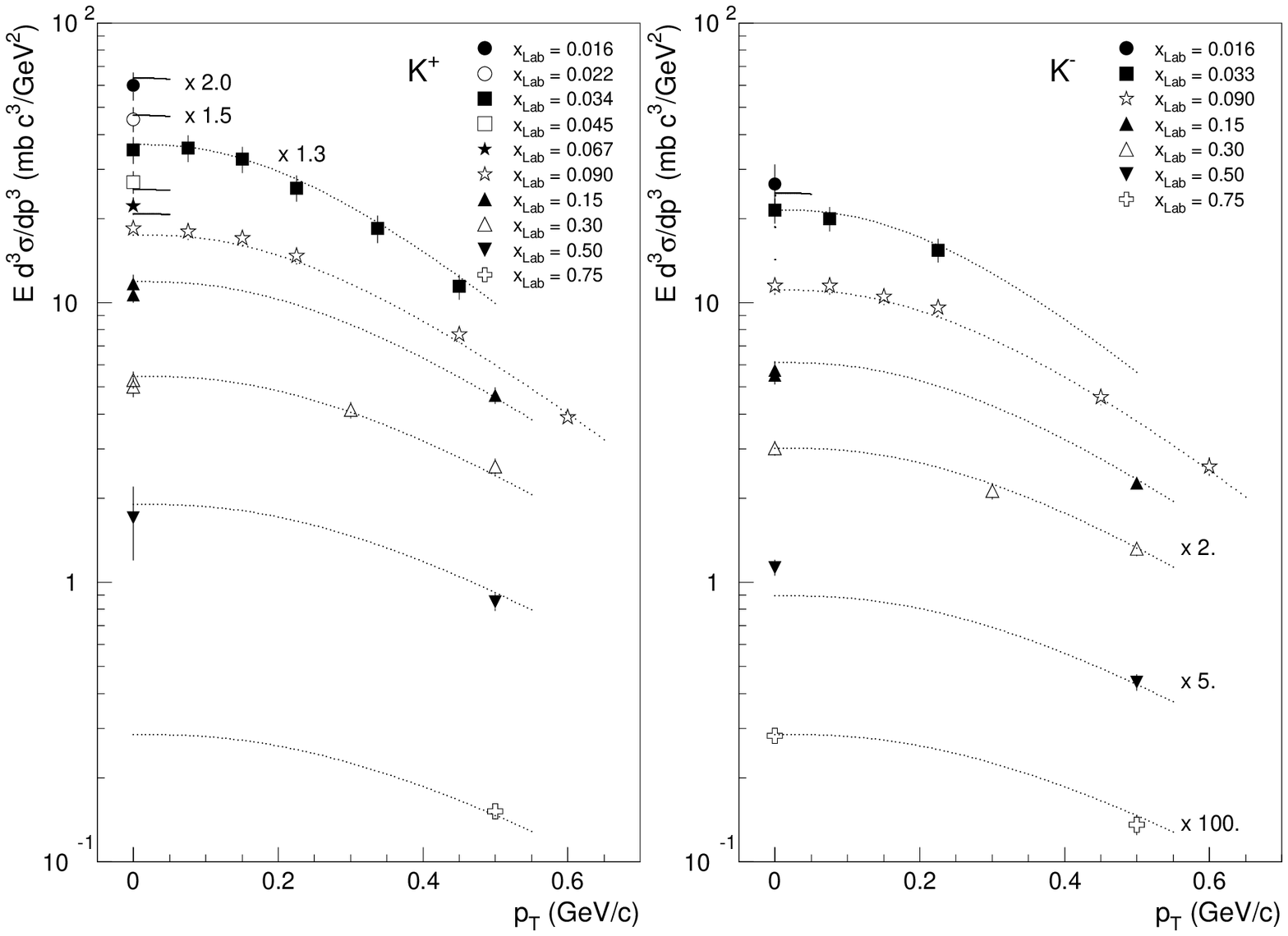,width=0.8\linewidth}}
  \caption{Invariant cross section as a function of $p_T$: 
          (top-left)    positive pions; (top-right)    negative pions;
          (bottom-left) positive kaons; (bottom-right) negative kaons.  
          Data collected at the same $x_{Lab}=p/p_{inc}$, where $p$ 
          and $p_{inc}$ are the momenta of the detected particle and of 
          the incident proton in the laboratory reference frame, are 
          displayed with the same symbol.           
          The best-fit obtained with the parameterization described 
          in the text is superimposed.}
  \label{fig2}
  \end{center}
\end{figure*}

\begin{figure*}
  \begin{center}
  \mbox{\epsfig{file=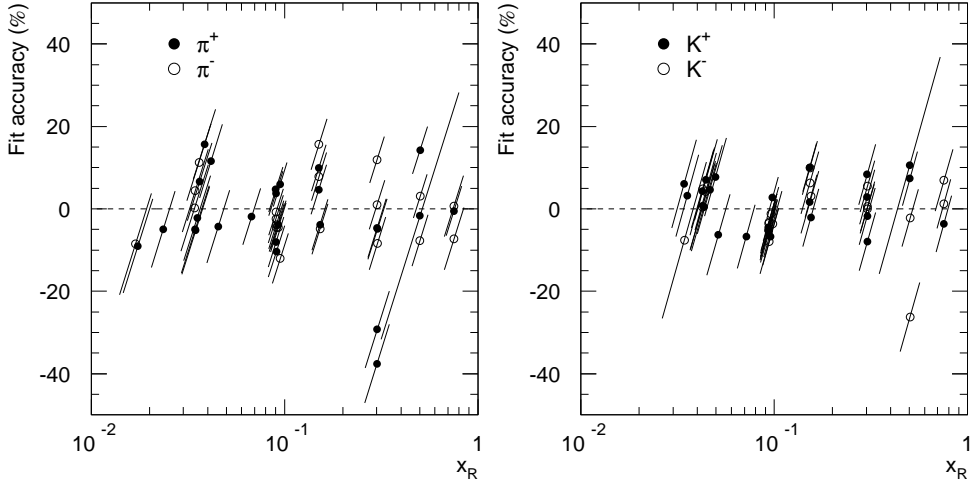,width=0.8\linewidth}}
  \caption{Percent difference between the best-fit prediction of 
           the parameterization proposed in this work and experimental 
           data (fit accuracy) for pions (left) and kaons (right) as
           a function of $x_R$.} 
  \label{fig3}
  \end{center}
\end{figure*}

The comparison between the empirical parameterization and the
experimental data is shown in figure \ref{fig2}. The accuracy of the
parameterization of the $\pi^{\pm}$ and $K^{\pm}$ data is displayed in
figure \ref{fig3}, showing the relative discrepancy between our
parameterization and the experimental data as a function of $x_R$. 
The proposed parameterization gives an accurate fit of $K^{\pm}$ data
with a reduced $\chi^2 \simeq 0.85$, while the reduced $\chi^2$ is
somewhat larger ($\chi^2/N_{dof} = 77.1/37$) for $\pi^{\pm}$ data. 
This partly reflects our difficulty to parameterize the $x$ dependence 
of the $p_T$ distribution, although about 1/3 of the $\chi^2$ is 
contributed by the two data points (one measured by NA20 and 
the other by NA56/SPY) for positive pion production in the forward 
direction at $x=0.3$, which are about 30-40\% off the best-fit 
prediction. 
A reduced $\chi^2$ around 1 is obtained, if a relative error 
of 10\% on each experimental point is assumed. We conclude that the 
proposed empirical formulae are adequate to describe NA20 and NA56/SPY 
data with a 10\% accuracy.

Proton and anti-proton data, which are of less direct interest for 
neutrino beams, have been parameterized with the following empirical 
formulae:
\begin{eqnarray}
\label{eqprot}
\nonumber
(E \times \frac{d^3 \sigma}{d p^3})_{p Be \to p X}  
\nonumber    & = & A (1+Bx_R) (1-x_R)^{b p_T^2} \times \\
             &   & (1+ap_T + \frac{a^2}{2} p_T^2) e^{-ap_T}
\end{eqnarray}
\begin{eqnarray}
\label{eqapro}
\nonumber
(E \times \frac{d^3 \sigma}{d p^3})_{p Be \to \overline{p} X} 
\nonumber     & = & A (1-x_R)^\alpha x_R^{-\beta} \times \\
              &   & (1+ap_T + \frac{a^2}{2} p_T^2) e^{-ap_T}
\end{eqnarray}
For anti-protons a functional similar to the one given in (\ref{eq1}) 
has been adopted, except that an exact factorization in $x_R$ and $p_T$
has been assumed, since this was sufficient to give a reasonable fit 
to data. For protons the ``leading particle effect'' had to be taken 
into account. A reasonable fit to data has been obtained by following 
the empirical observation that the longitudinal momentum distributions
of the leading nucleon in $pp$ collisions is flat \cite{Coc71}, which 
translates into a linear rise of the inclusive invariant cross section 
as a function of $x_R$. The transverse momentum distribution is also 
affected by the leading particle effect, resulting in an enhanced 
leading particle production in the forward direction (see for example 
\cite{Tay76,Allab}). In the proton fit, this is empirically accounted 
for by the term $(1-x_R)^{b p_T^2}$. 
The parameter $a$, that controls the shape of the $p_T$ distribution 
for non-leading particle production, has been assumed to be the same 
for protons and anti-protons. 

\begin{figure*}
\begin{center}
\mbox{\epsfig{file=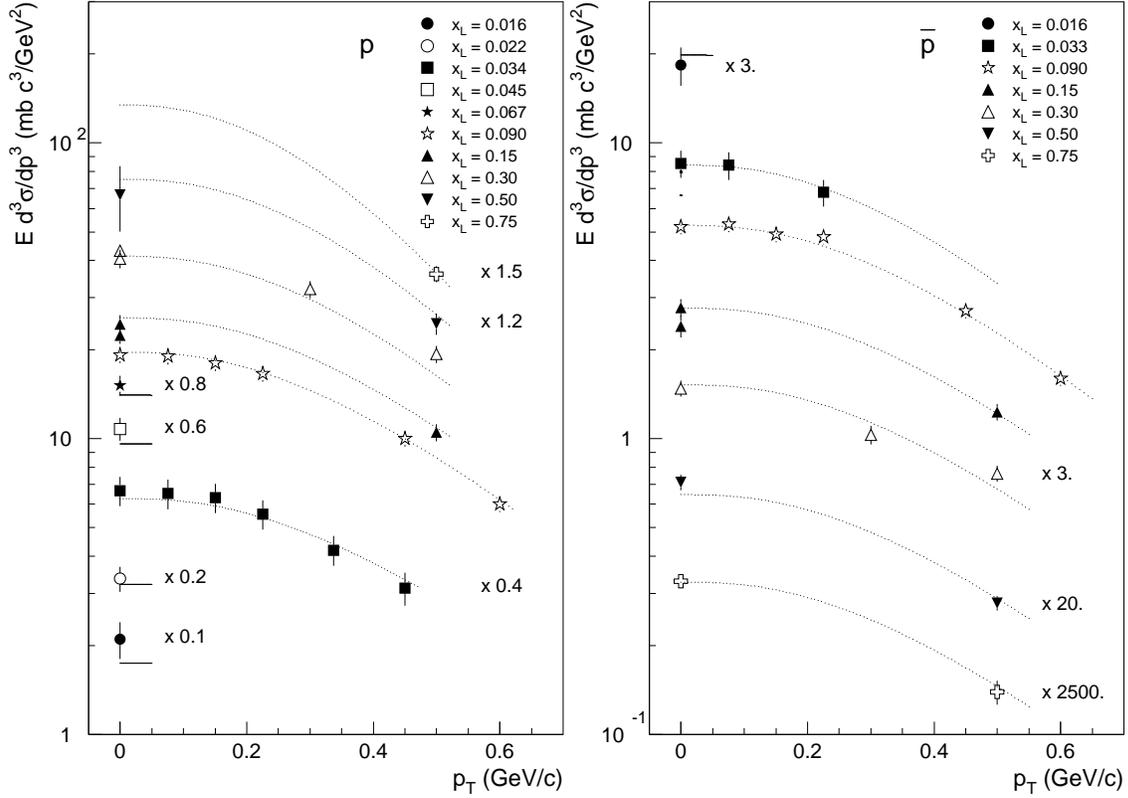,width=0.85\linewidth}}
\caption{Invariant cross section as  a function of $p_T$: (a) protons;
(b)  anti-protons.  Data collected  at  the same  $x_{Lab}=p/p_{inc}$,
where $p$ and  $p_{inc}$ are the momenta of  the detected particle and
of the incident proton in the laboratory reference frame are displayed
with the same symbol.  The empirical parameterization described in the
text is superimposed.}
\label{figp}
\end{center}
\end{figure*}

Results of these fits are also given in Table~\ref{tab:tabone} and 
the comparison between the empirical parameterization and the 
experimental data is shown in figure \ref{figp}. Our parameterization
gives a satisfactory description of proton and anti-proton inclusive
production, with a reduced $\chi^2$ about 1, in the range covered by
NA56/SPY and NA20 data. At larger values of $x$, proton production by 
means of diffraction should occur. This is not described by our 
parameterization. 

\section{Production of neutral Kaons}

The knowledge of neutral kaon production in p-Be interactions is
important for the exact   calculation of the $\nu_e$ background in
neutrino beams coming from  $K^0_L$ decays. As an example, in the case
of the WANF beam serving the CHORUS and NOMAD experiments, the estimated 
contribution to the $\nu_e$ content of the beam due to $K^0_L$ is
around 15\%. 
In addition, the knowledge of the neutral kaon
production is also of interest for the  experiments on CP-violation
using neutral kaon beams and it is particularly relevant in searches
for rare neutral kaon decays. 

A rather complete set of data on $K_S^0$ production at angles
different from zero was collected by Skubic et al. \cite{Sku78},
using a 300 GeV/c proton beam hitting Be targets. Measurements of
$K^0_S$ production in the forward direction from 200 GeV/c protons on
beryllium were performed by Edwards et al. \cite{Edw78}. Both these
experiments, however, covered a region of secondary momenta
corresponding to $x_F > 0.25$, which only partially overlaps to the
momentum region important for present and future neutrino beams.  

Additional information on neutral kaon production can be obtained
starting from the available  measurements of charged kaon production
in p-Be interactions. Charged and neutral kaon production rates 
can be related in the frame of a simple quark parton model, in
which valence quarks ($q_v$) and sea quarks ($q_s$) are considered.

In the simplest model, isospin symmetry may be assumed, giving
$u_v/d_v = 2$, $u_s = \overline{u}_s = d_s = \overline{d}_s$ and  
$s_s = \overline{s}_s$, from which one finds that the production of 
charged and neutral kaons should be related by
\begin{equation}
\label{eq34}
N(K^0_S)=N(K^0_L)=\frac{1}{4}(N_{K^+}+3 \cdot N_{K^-})
\end{equation}
This model has been shown to provide an accurate description of the
asymmetry of $K^0$/$\overline{K^0}$ production in the range
$0.18<x_F<0.36$ in neutral kaons beam-lines \cite{Dob90}. We have
verified that it also agrees within 15\% with direct measurements 
of $K_s$ production up to around $x_F=0.5$. At larger values 
of $x_F$, i.e. in the fragmentation region of the proton, we find 
that a reasonable description of existing $K_s$ production data 
requires a ``dynamic'' model in which the $u_v/d_v$ ratio is let to be 
$x$-dependent and estimated from the  $\pi^+/\pi^-$ production
ratio, following the arguments given by Ochs \cite{Och77}, who has
demonstrated a remarkable empirical similarity between the $\pi^+ /
\pi^-$ production ratio in $pp$ collisions and the $u/d$ quark ratio
measured in deep-inelastic lepton-proton scattering. The exploitation
of this model is beyond the aim of this paper\footnote{A similar
``dynamic'' prediction of neutral kaon production was already
considered in ref. \cite{Bor96}. These models however seems less
accurate than formula (\ref{eq34}) in the description of the
$K^0$/$\overline{K^0}$ production asymmetry at small/intermediate
$x_F$.} and neutral kaon production is always estimated from eq. 
(\ref{eq34}) in the following.

\section{Particle production for neutrino beams} 

\subsection{Scaling to targets of different materials} 

Beryllium targets have been used in most neutrino beams derived from
proton beams extracted from  the Super Proton Synchrotron (SPS) at CERN, 
but the use of graphite as target material seems promising in view 
of operation with short spills in fast extracted proton beams.
In particular, the use of a graphite target is foreseen in the design
of both the CNGS and the NuMI beams \cite{CNGS,NUMI}. Prescriptions 
to rescale the inclusive invariant cross-sections to different target 
materials are given in the following. 

Invariant cross sections $E  \frac{d^{3}\sigma^{hA}}{d p^{3}}$ for
hadron-nucleus interactions ($ h A \mapsto h' X$) depend on the
mass number $A$ of the target nucleus, via parameterizations of the
type:   
\begin{equation}
 E  \frac{d^3 \sigma^{hA_{1}}}{d p^3} = (\frac{A_1}{A_2})^{\alpha} \cdot
 E  \frac{d^3 \sigma^{hA_{2}}}{d p^3}  \label{eq2}
\end{equation}
where a value for $\alpha = 2/3$ would correspond to the case where 
particle production off a nucleus is identical to the production 
off a single nucleon per inelastic collision.  
In accordance with the scaling hypothesis, $\alpha$ has been found 
to be weakly dependent on the incident beam momentum. It depends on 
the incident hadron type $h$ and it is a smooth function of $p_T$ 
and $x_F$ of the produced hadron. It has moreover been experimentally 
observed that, in first approximation, it is independent of the 
detected particle type, with perhaps the exception of anti-protons 
\cite{Bar87}. 

A parameterization of $\alpha$ as a function of $x_F$ has been 
proposed by Barton et al. on the basis of their and previous data 
at $p_T=0.3$ GeV/c \cite{Bar87}. A $p_T$ dependence of $\alpha$ 
has been clearly observed in $K^0_S$ and  $\Lambda$ production 
by Skubic et al. \cite{Sku78}. 
A suitable representation of the whole set of data can be obtained 
with the parameterization:
\begin{equation}
\alpha(x_F)=(0.74 -0.55 \cdot x_F + 0.26 \cdot x_F^2) \cdot
(0.98 + 0.21 \cdot p_T^2 ) 
\label{eq3}
\end{equation}
where the $x_F$ dependence is taken from the fit of Barton et al. at 
$p_T=0.3$ GeV/c and the $p_T$ dependence is fitted to Skubic et al. 
data and normalized in such a way that it reduces to the 
parameterization of Barton et al. at $p_T=0.3$ GeV/c. 

In conclusion, a conservative estimate of the uncertainty in the
extrapolation from  beryllium to carbon data sits (in the $p_T$ range
of interest for neutrino beams: up to $\sim 600$ MeV/c) around 5\%, 
on the top of a measurement error of 5--10\%, depending on the
secondary momentum, for the cross sections on beryllium. The estimate 
of this systematic uncertainty is based both on data collected by 
Barton et al. \cite{Bar87} and on the extensive compilation of J. Kuhn 
on nuclear dependence for $p A \to \pi^{-} X$ interactions~\cite{Kuh78}.

\subsection{Scaling to different centre of mass energies}

The NuMI neutrino beam at FNAL is planned to be derived from a primary
proton beam of 120 GeV/c momentum, resulting in a centre-of-mass
energy about two times smaller than that available at NA56/SPY and NA20.  
Besides its general interest, a test of the scaling hypothesis 
of one-particle inclusive invariant cross-sections is thus relevant 
to assess the range of validity of the proposed parameterization. 
This has been studied by comparing our parameterisation to available 
$pA$ data collected at different centre-of-mass energies. 

\begin{figure*}
\begin{center}
\mbox{\epsfig{file=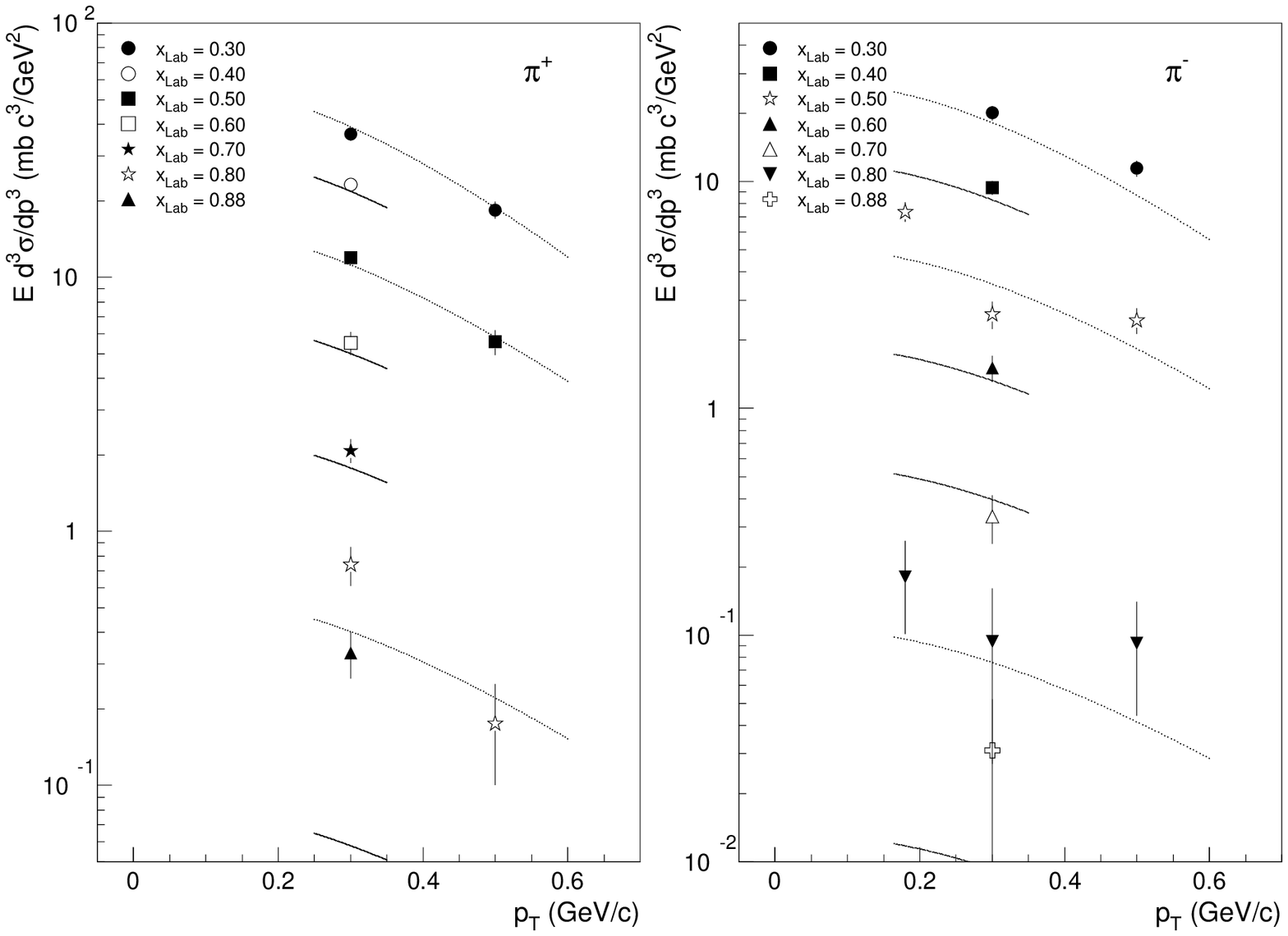,width=0.75\linewidth}}
\end{center}
\caption{Comparison between the proposed parameterization of 
NA56/SPY \cite{SPY2} and NA20 \cite{NA20} data (line) and the one-pion 
inclusive invariant cross sections in $pC$ interactions at 100 GeV 
as measured by \cite{Bar87}. Positive (negative) pions are shown 
in the left (right) panel.}
\label{Barfig}
\begin{center}
\mbox{\epsfig{file=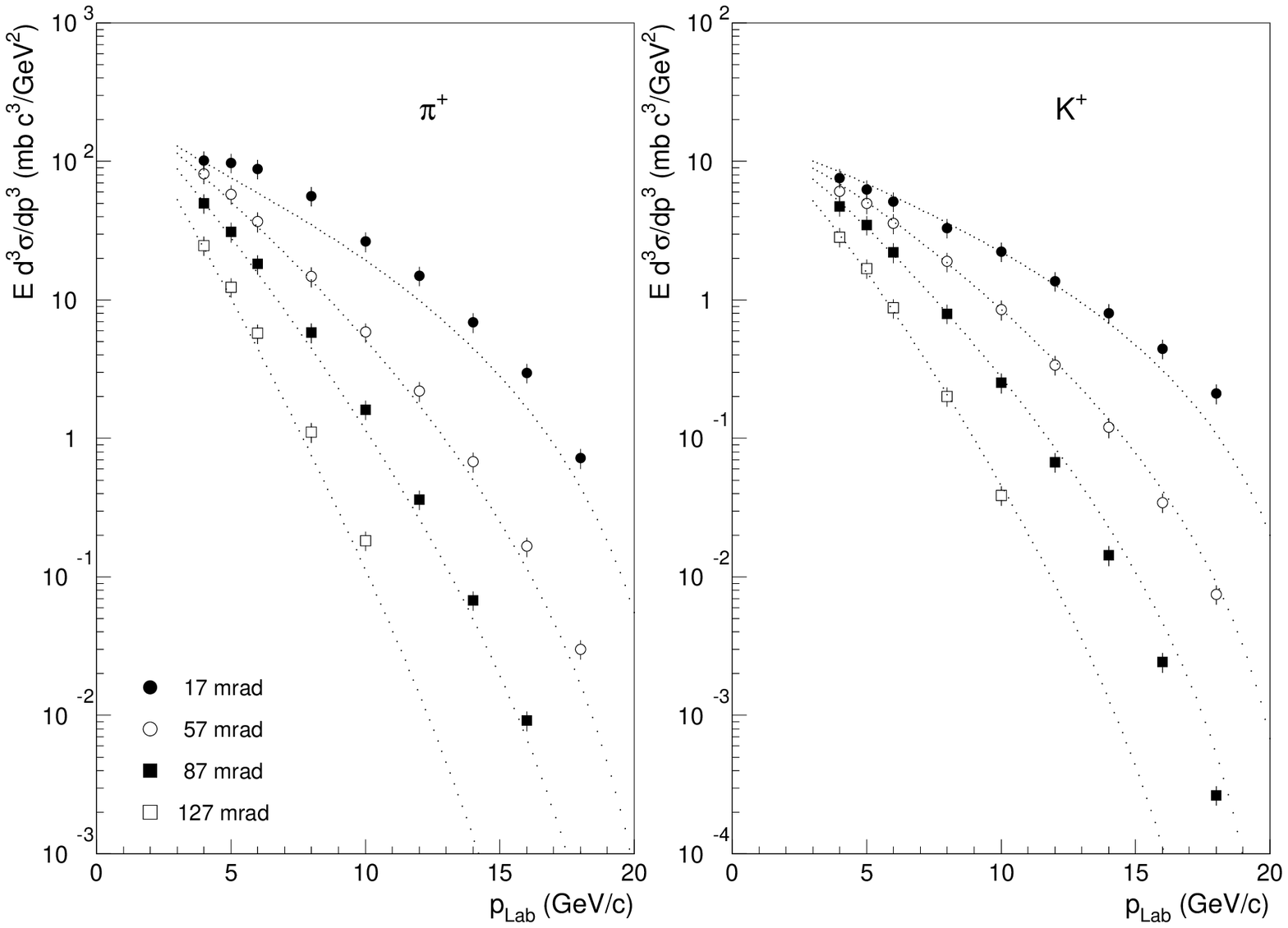,width=0.75\linewidth}}
\end{center}
\caption{Comparison between the proposed parameterization of 
NA56/SPY \cite{SPY2} and NA20 \cite{NA20} data (dotted line) and the 
one-pion (left) and one-kaon (right) inclusive invariant cross sections 
in $pBe$ interactions at 24 GeV as measured by \cite{Eic72}.}
\label{Eicfig}
\end{figure*}

In figure \ref{Barfig}, our prediction for one-particle inclusive 
invariant cross section are compared to data collected with 100 GeV/c 
protons on carbon target \cite{Bar87}, where the extrapolation from 
Be to C has been made using formulae (\ref{eq2}) and (\ref{eq3}). 
As discussed above  $x_R = E^*/E^*_{max}$ has been used as scaling 
variable. The agreement between pion data shown in the figure and 
our empirical parameterization is excellent up to about $x \sim 0.8$. 
A good agreement is also found when kaon data are considered, although 
the precision of kaon data from ref. \cite{Bar87} is poorer.

Figure \ref{Eicfig} shows a comparison of our parameterization with 
one-particle inclusive production data measured with 24 GeV/c protons 
hitting a Beryllium target \cite{Eic72}. In that work, Lorentz 
invariant particle densities $\omega(p_{Lab},\theta) $ were measured 
as a function of the particle momentum $p_{Lab}$ and production angle 
$\theta$. These data have been converted into invariant cross sections 
using the relation 
\begin{equation}
E \times \frac{d^3\sigma}{dp^3} = \frac{1}{2} \sigma_{abs} 
                                              \omega(p_{Lab},\theta)
\end{equation}
where $\sigma_{abs}$ is the absorption cross section for $pBe$ 
\cite{Car79}. A reasonable agreement is observed in the shape of 
the distributions for $\pi^+$ and $K^+$, although the estimated 
production of $\pi^+$ is about 35 $\pm$ 15\% lower than that measured. 
This is also true for negative pions, not displayed in the figure, 
while the agreement is somewhat worse for the other particles. 



Given its general interest, we have tested the scaling hypothesis on
the same data also using $x_F$ and $x_{Lab} = p/p_{inc}$, the latter 
defined as the ratio of the momentum of the  
detected particle in the laboratory reference frame to the momentum of
the incoming proton, as longitudinal variables. A better agreement 
with $\pi^\pm$  data is obtained, if $x_{Lab}$ is used, which however
shows a worse agreement to kaon data\footnote{All the variables are off 
roughly by the same amount in the description of the $K/\pi$ production 
ratio.}. The variable $ \Delta y = y_{max} - y$, where 
$y=\frac{1}{2}\log((E+p_L)/(E-p_L))$ is the rapidity of the 
produced particle and $y_{max}$ is the maximum rapidity kinematically 
available to that particle, has also been tried. At variance with the
previuos longitudinal variables, $\Delta y$ is Lorentz invariant and 
scales the phase space at different centre-of-mass energies
independently of the reference frame. 
Although we find a somewhat improved agreement to data in this case,
scaling to data collected with 24 GeV/c protons is only approximate
even with this variable.

Still, in the centre-of-mass energy range of interest for present 
high-energy neutrino beams, the agreement between our parameterization 
scaled according to $x_R$ and the data is satisfactory. 

\subsection{Yields from finite length targets} 
\label{yields}

With reference to eq. (\ref{I2Y}), the differential particle production
along the target can be parameterized as: 
\begin{equation}
\frac{dY(E,p_T,z)}{dz} = \frac{N_0 \rho \lambda_p}{100~A} 
                         \frac{p^3}{E} 
                         \left( E \times \frac{d^3 \sigma}{d p^3}\right)
                         f(z)
\end{equation}
where $f(z)dz$ is the probability that the outgoing particle be
produced at a depth $z$ to $z+dz$ inside the target and the other
quantities have been introduced in Section \ref{ics}.

In general $f(z)$ will also depend on the production angle of the 
secondary particle \cite{Bon96}. In the naive reabsorption model
introduced in Section \ref{ics}, one has:
\begin{equation}
\label{secca}
f(z,\theta) = \frac{1}{\lambda_p} e^{-z/\lambda_p} 
                     e^{-z'(z,\theta)/\lambda_s}
\end{equation}
which can be obtained by differentiating the target efficiency 
$F(L)$ of equation (\ref{targeff}), with $z'(z,\theta)$ representing 
the target length that the produced hadron has to cross to escape 
from the target ($L-z$ for forward production). 
This naive model is only adequate to give a first order description of
particle production from relatively thin targets or at large $x$, but
in general particle production by means of cascade processes cannot 
be neglected.  

Since all types of hadrons can be produced when an energetic hadron of
any flavour interacts in the target, the production of particles
from a target of finite length is governed by a set of coupled
transport equations that depend on the properties of the particles,
their interactions and on the structure (geometry and material) of the
target. Given their complexity, the solution of these equations is in 
general addressed numerically with hadron cascade/transport codes,
calibrated on a large set of hadron production data.
However, in the spirit of this work, the main features of the cascade 
process in needle-shaped targets, like those used for neutrino beams, 
can be described in a simple parametric form and tuned on NA56/SPY 
data on yields from targets of different length and geometry.

In hadronic interactions, about half of the available energy is
typically dissipated in multiple particle production  
(inelasticity $k \simeq 0.5$) with average 
transverse momentum $\langle p_T \rangle \simeq 0.35$ GeV/c . 
The remainder of the energy is carried 
by fast forward-going ``leading'' particles. Only these particles 
are responsible for the propagation of the cascade in thin and long
targets, since they are almost collinear with the primary 
beam\footnote{The divergence of leading particles with respect to the 
primary beam is typically $\langle p_T \rangle/(1-k) p_{inc} \simeq $ 
few mrad for proton energies of order a few hundred GeV.} and  their energy 
is large enough to result in a sizeable yield of additional energetic 
particles. 

In the collinear approximation and assuming that leading particles 
in $pA$ interactions are mostly protons, thus characterized by an
effective mean free path $\lambda_p$ equal to that of primary protons, 
the naive reabsorption model of equation (\ref{secca}) can be improved 
by the expansion: 
\begin{equation}
\label{collin}
f(z) = \frac{1}{\lambda_p} e^{-z/\lambda_p} 
       [ 1 + A_h(x) \frac{z}{\lambda_p} ] 
        e^{-z'(z,\theta)/\lambda_s}     
\end{equation}
where the second term in brackets accounts for particle production
by reinteractions of secondary particles (tertiary particle
production), while higher rank contributions are neglected, since
neutrino targets are typically of order $2 \lambda_p$ and the less
energetic is the reinteracting particle the lower is the yield of
produced particles.  

In equation (\ref{collin}), $A_h(x)$ weights the probability that 
reinteractions of secondary particles will result in a produced
hadron $h$ of fractional longitudinal momentum $x$ and is given by:
\begin{equation}
\nonumber
A_h(x) = \sum_{h'} \int_x^1 dx' G_{h' \to h}(x,x') \Phi_{h'} (x') 
\end{equation} 
where $G_{h' \to h}(x,x')$ is the $p_T$ integrated cross section for
inclusive production of $h$ in $h'$ interactions and $\Phi_{h'} (x')$
is the flux of $h'$ particles. Thus, $A_h(x)$ is a function of $x$ 
and $h$ only. Its dependence on these variables has been derived 
from NA56/SPY measurements of inclusive yields in the forward direction 
from finite length targets.

\begin{figure}
\begin{center}
\mbox{\epsfig{file=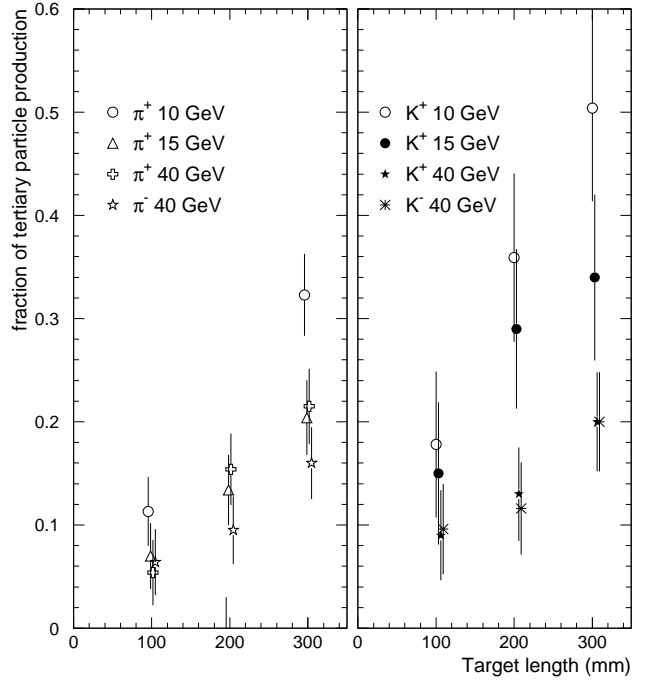,width=\linewidth}}
\end{center}
\caption{Fraction of tertiary particle production as a function of 
the target length as derived from published NA56/SPY data \cite{SPY2}.} 
\label{terlen}
\end{figure}

In the short target approximation, valid until the target length is 
$L << \lambda_p \lambda_s / (\lambda_s - \lambda_p)$, integration of 
equation (\ref{collin}) in the forward direction (i.e. $z'=L-z$) 
predicts the fraction $t(L)$ of tertiary particle production to be
approximately a linear function of $L$ given by:
\begin{equation}
\label{linear}
t(L) \simeq A_h(x) \frac{L}{2 \lambda_p}.
\end{equation}
This linear behaviour is in agreement  with experimental data, 
as shown figure \ref{terlen}, where the fraction of tertiary
production as a function of target length for $\pi^{\pm}$ and 
$K^{\pm}$ as derived from NA56/SPY data is displayed, and was exploited 
in section \ref{ics} to derive inclusive invariant cross sections. 

\begin{figure}
\begin{center}
\mbox{\epsfig{file=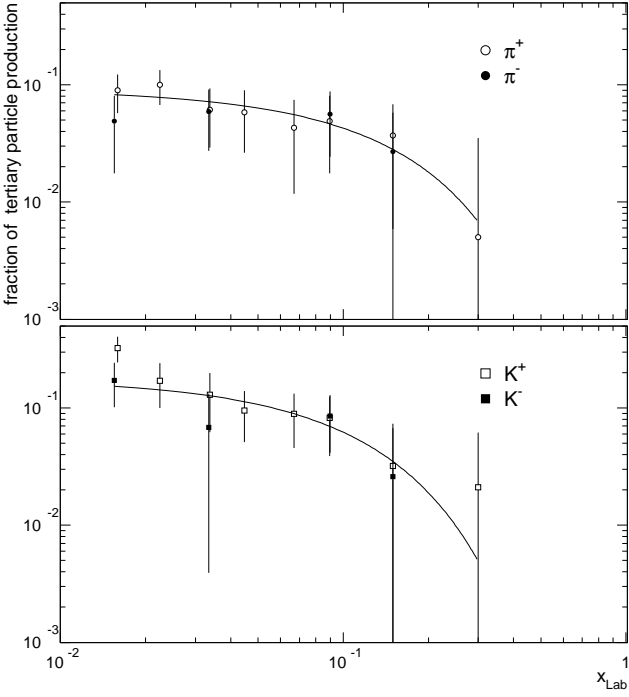,width=\linewidth}}
\end{center}
\caption{Fraction of tertiary particle production in a 100~mm Be 
target as a function of $x_{Lab}$. Data are derived form published 
NA56/SPY data \cite{SPY2} and the curves represent the best-fit described 
in the text.}
\label{termom}
\end{figure}

Compared to the naive absorption model, the excess of particle 
production present in the NA56/SPY data increases with decreasing
momentum. Figure \ref{termom} shows the fraction of tertiary particles
produced in the forward direction for a 100~mm target as a function 
of the fractional longitudinal momentum $x_{Lab} = p_{Lab}/p_{inc}$. 
The curves shown in the figure represent the best-fit of $A_h(x)$ to
data according to the empirical parameterization: 
\begin{equation}
\label{terzeq}
A_h(x_{Lab}) = A^{ter}_h (1-x_{Lab})^{b^{ter}_h}
\end{equation}
where a common value of $A_h^{ter}$ and $b_h^{ter}$ has been chosen for
positive and negative particles. The best-fit results are reported in 
table \ref{tab:terfit}. 

\begin{table}
\begin{center} 
\begin{tabular}{cccc} 
\hline\noalign{\smallskip}
 $A^{ter}_{\pi}$ & $A^{ter}_K$  & $b^{ter}_{\pi}$ & $b^{ter}_K$ \\ 
\noalign{\smallskip}\hline\noalign{\smallskip}
     0.80        &   1.56       &   7.3           &   10.1      \\ 
\noalign{\smallskip}\hline
\end{tabular} 
\caption{Best-fit values for parameters of equation (\ref{terzeq}) 
to NA56/SPY data (see text for details).} 
\label{tab:terfit}  
\end{center} 
\end{table}

In this simple model, no $p_T$ dependence of tertiary particle 
production is predicted. Indeed, assuming Feynman scaling and
the (approximate) factorization in $x$ and $p_T$ of the invariant
cross-section, the $p_T$ distribution of tertiary particles is
expected to be the same as that of secondary particles. It follows 
that the divergence of leading particles with respect to the primary 
beam direction is in general negligible as compared to the typical 
production angle of lower momentum tertiary particles (
collinear approximation).
These considerations are supported by NA56/SPY data that show no $p_T$ 
dependence of the transverse momentum distribution of tertiary
particle production in a 300~mm Be needle-shaped target once 
mere geometric effects are considered (see figure 20 of Ref. \cite{SPY2}). 

Equation (\ref{collin}) is valid for NA56/SPY and NA20 targets, but fails  
to give the correct prediction for tertiary particle production when
the target length is of order $\lambda_{geom} \equiv R \cdot p_l / 
\langle p_T \rangle$, where $R$ is the transverse dimension (radius) 
of the target and $p_{l}=(1-k) \cdot p_{inc}$ is the typical momentum 
of the leading particle. 
In this limit, there is a sizeable probability that some of
the leading particles also escape from the side of the target before 
interacting. This can be accounted for, modifying  equation
(\ref{collin}) into: 
\begin{equation}
\label{geom}
f(z) = \frac{ e^{-z/\lambda_p} }{\lambda_p} 
       \left[ 1 + A_h(x) \int_0^z \frac{dy}{\lambda_p} w_z(y) \right] 
              e^{-z'(z,\theta)/\lambda_s}     
\end{equation}
where $w_z(y)$ is an acceptance factor, which in general depends on 
the $p_T$ and energy distributions of the leading particles and on 
the target geometry\footnote{In equation (\ref{geom}), the collinear
approximation has been again advocated to factorize the acceptance
factor and the reinteraction probability of leading particles.}.
For cylindrical symmetry, one can write:
\begin{equation}
\label{gggg}
w_z(y)=\frac{1}{\langle p_T \rangle} 
                \int_0^{p_T^{max}} h(p_T) p_T dp_T
\end{equation}
where $p_T^{max} \simeq p_l R / (z-y)$ in the collinear approximation and
$h(p_T)$ is the $p_T$ distribution of the produced particles.


In the simulation of neutrino beams, discussed in the following, the
computation of this acceptance factor has been performed under the
approximation that leading particles carry on average half of the
primary beam momentum and that their transverse momentum distribution
is represented by a pure exponential of average transverse momentum
$\langle p_T \rangle$. For cylindrical symmetry this allows the
analytical integration of equations (\ref{geom}) and (\ref{gggg}).
In case of continuous targets of radius $R$, the integration yields:
\begin{equation}
\label{gappr}
f(z) = \frac{ e^{-z/\lambda_p} }{\lambda_p} 
       \left[ 1 + A_h(x) \frac{z}{\lambda_p} (1-e^{-\lambda_{geom}/z})
       \right] 
              e^{-z'(z,\theta)/\lambda_s},     
\end{equation}
which reduces to equation (\ref{collin}), predicting a linear increase 
of the differential production of tertiary particles along the target, 
until $z << \lambda_{geom}$ (short target approximation). At larger 
depth inside the target, the differential production of tertiary 
particles will tend to saturate to a value proportional to 
$\lambda_{geom}/\lambda_p$. 
 
In case of a segmented target, made of several rods interleaved by 
air gaps, as that used in the WANF beam and considered for the 
forthcoming CNGS beam at CERN, the details of the target geometry 
have to be considered. The analytical integration of equations 
(\ref{geom}) and (\ref{gggg}) is still possible under the same 
approximation as that adopted to derive equation (\ref{gappr}) 
and it has been used in the evaluation of the efficiency of 
segmented targets.

\begin{figure*}
\begin{center}
\mbox{\epsfig{file=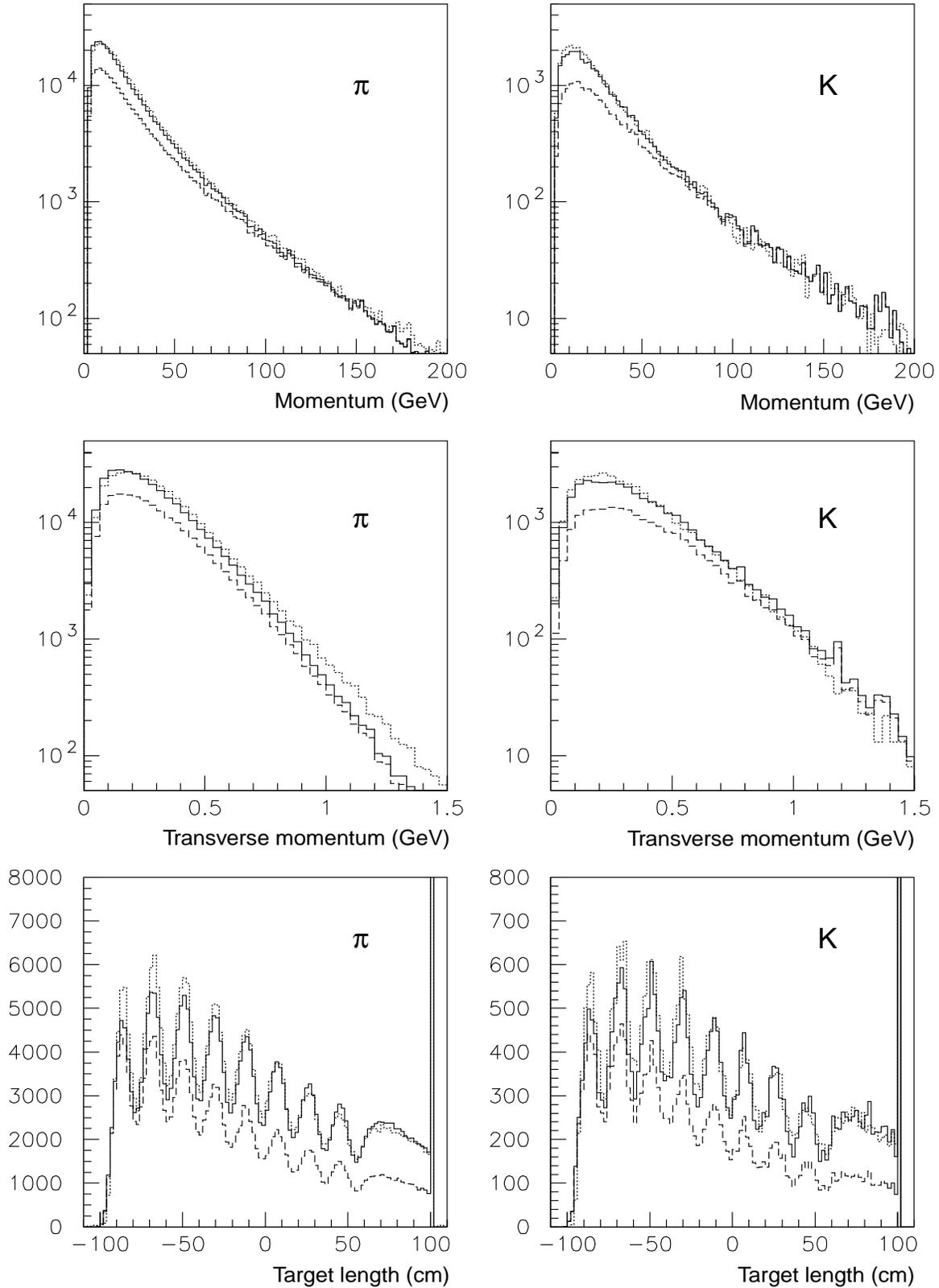,width=0.85\linewidth,height=0.9\textheight}}
\end{center}
\caption{Longitudinal (upper panels) and transverse (middle panels) 
momentum distributions of secondary mesons from the CNGS target 
(within 30 mrad acceptance). On the lower panels the exit points
 of the secondary mesons from the CNGS target (within 30 mrad
acceptance) are shown. The dotted line is the FLUKA prediction, 
the dashed line is our computation without tertiary contribution, 
the solid line is our prediction with tertiary mesons.}
\label{figyield}
\end{figure*}

In Figure~\ref{figyield} we present predictions for secondary meson 
production at target level, based on our parameterization of the
inclusive invariant cross-sections corrected for target efficiency. 
Momentum spectra (top), angular distributions (middle) and exit point 
profiles (bottom) of the secondary mesons in the case of the CNGS
target geometry and for an angular acceptance of 30 mrad (much larger 
than the horn aperture of the CNGS beam-line described in the next 
Section) are shown. For sake of comparison, the predictions of a 
FLUKA based full Monte Carlo program \cite{FONE,CNGS} are also 
shown. The agreement is remarkable, provided that tertiary 
particle production is taken into account in the evaluation of 
the target efficiency, as discussed in this Section.

A slight disagreement in the $p_T$ distributions of all charged 
particles is visible mainly at large $p_T$, that translates in a 
slightly more forward peaked distribution of the exit points along 
the target. This disagreement appears in a kinematic region not 
covered by NA56/SPY and NA20 data. However, as remarked in the next
Section, neutrino beam predictions are only slightly affected by 
this large $p_T$ region, since it falls outside the momentum 
and angular acceptance selected by the focusing optics. The systematic
difference between the two models is thus negligible. 

\section{Simulation of neutrino beams}

As it is well known, a ``classic'' wide band neutrino beam is produced
from  the decay of mesons, mostly $\pi$'s and $K$'s. Mesons are
created by the  interaction of a proton beam into a needle shaped
target, they are  sign-selected and focused in the forward direction
by two large acceptance  magnetic coaxial lenses,  conventionally called 
at CERN horn and reflector, and finally they are let to decay into an 
evacuated tunnel pointing toward the detector position.

In case of positive charge selection, the beam content is mostly
$\nu_\mu$  from the decay of $\pi^+$ and $K^+$.  Small contaminations
of $\overline \nu_\mu$ (from the defocused $\pi^-$ and $K^-$) and $\nu_e$
(from three-body  decay of $K$'s and $\mu$'s) are present at the level
of few percent.

\begin{figure*}
\begin{center}
\mbox{\epsfig{file=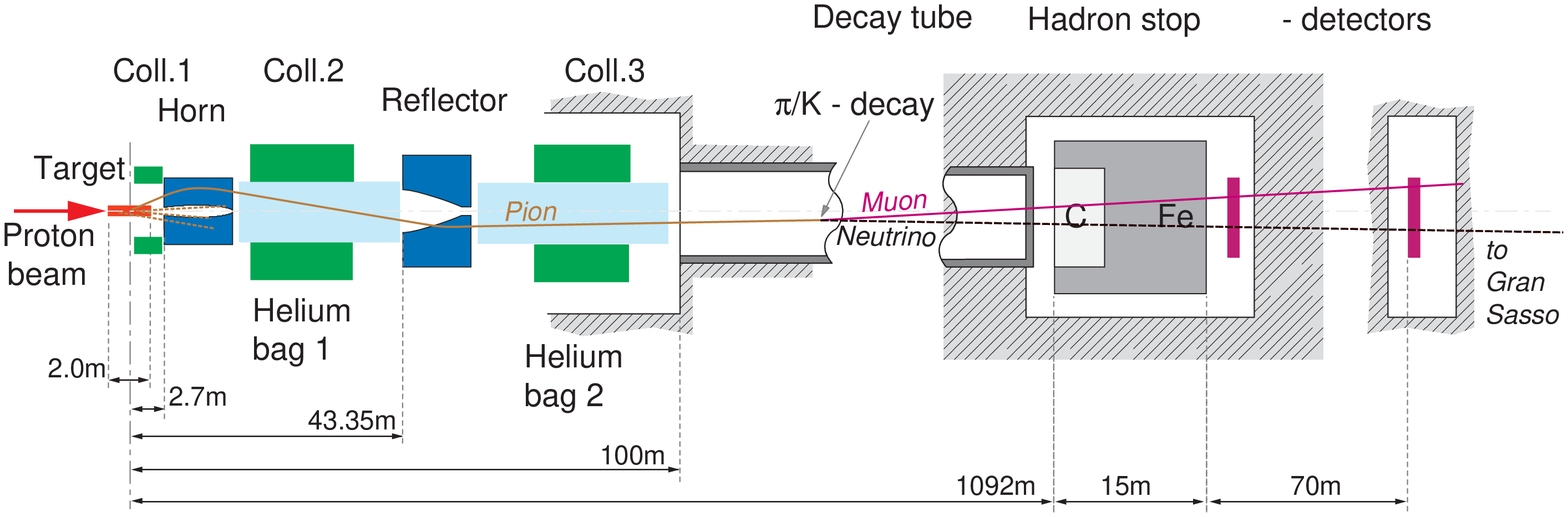,width=0.85\linewidth}}
\end{center}
\caption{Schematic layout of the future CNGS neutrino beam line.}
\label{fig:cngslay}
\end{figure*}

As an example, a schematic layout of the future CERN to Gran Sasso
neutrino  beam, CNGS \cite{CNGS}, is shown in Figure~\ref{fig:cngslay}. 
Its main characteristics, relevant for beam simulation purposes, 
are listed in Table~\ref{tab:ngstab} together with those of the old
West Area Neutrino Facility (WANF) at CERN in the configuration set up 
for the CHARM II experiment\cite{WANF}.

\begin{table*}[htb]
\begin{center}
\begin{tabular}{lcc}
\hline\noalign{\smallskip}
Beam line                  & WANF                & CNGS     \\     
\noalign{\smallskip}\hline\noalign{\smallskip}
Target material            & beryllium           & graphite \\
Target rod length          & 10 cm               & 10 cm    \\
Target rod diameter        & 3 mm                & 4 mm     \\
Number of rods             & 10                  & 8        \\
Rod separation             & 9 cm                & 9 cm     \\
Additional end--rod length & --                  & 50 cm    \\
\noalign{\smallskip}\hline\noalign{\smallskip}
Proton energy           & 450 GeV             & 400 GeV             \\
Proton beam focal point & \multicolumn{2}{c}{50 cm from start of target}\\
Expected pot/year:      & \hfill              & \hfill              \\
in shared SPS mode      &  $<2\times 10^{19}$ & $4.5 \times 10^{19}$\\ 
in dedicated SPS mode   & \hfill              & $7.6 \times 10^{19}$\\
\noalign{\smallskip}\hline\noalign{\smallskip}
Horn \& Reflector focusing momenta & 70--70 GeV  & 35--50 GeV   \\
Horn \& Reflector length            & 6.56 m      & 6.65 m      \\
Horn \& Reflector current           & 100--120 kA & 150--180 kA \\
Horn distance from focal point      & 11.5 m      & 2.7 m      \\
Refl.distance from focal point      & 82.8 m      & 43.4 m     \\
Horn acceptance                     & $\simeq$ 10 mrad & $\simeq$ 20 mrad \\
\noalign{\smallskip}\hline\noalign{\smallskip}
Decay tunnel length                 & 285 m       & 992 m    \\
Decay tunnel radius                 & 0.6 m       & 1.22 m   \\
Tunnel vertical slope               & +42 mrad    & -50 mrad \\
Pressure in decay tunnel            & 1 Torr      & 1 Torr   \\
\noalign{\smallskip}\hline
\end{tabular}
\caption{Main parameter list of the WANF (CHARM II set-up) and 
CNGS neutrino beam lines.}
\label{tab:ngstab}
\end{center}
\end{table*}

The neutrino fluxes for such a kind of beam are relatively easy to
predict  once the secondary meson  spectra are known, because the
meson decay kinematics is well understood and  the geometry of the
decay tunnel is quite simple.

Uncertainty in the estimation of the neutrino fluxes could arise
because  secondary mesons are selected over a wide momentum range
and over a wide angular acceptance ($\simeq 20$ mrad).

Re-interactions of secondary  mesons in the target and downstream
material contribute to reduce the  neutrino fluxes and increase the
uncertainty in the calculations (mainly for the wrong sign and wrong
flavour contaminations).  They are generally minimized  using a target
made of a number thin rods of low Z material interleaved  with empty
spaces (to let the secondary mesons exit the target without
traversing too much material). In addition the amount of material
downstream  of the target (i.e. horn and reflector conductor
thickness) is kept  to the minimum.

The parameterization of the secondary meson production from protons
onto a thin target, proposed in this paper, is thus well suited to be
used in neutrino  beam simulations both because it extends its
prediction over a wide range of longitudinal and transverse
momenta and also because  the small fraction of tertiary production 
from re-interactions in the target and downstream material can be 
accounted for with the approximations described in Section 
\ref{yields}.

A comparison of the neutrino flux prediction based on the proposed
parameterization with some measured spectra is thus an effective
estimator of the quality of the secondary mesons parameterization. 
For this purpose, our parameterization has been coupled with a 
neutrino beam simulation program able to provide rapid and accurate  
predictions of neutrino spectra at any distance (i.e. short and 
long base line). The comparison has been performed both with already 
published  data (CHARM II) and with predictions for the future CNGS 
long-baseline neutrino beam generated with GEANT and/or 
FLUKA based Monte Carlo programs.

\subsection{The simulation program}

The program is a stand-alone code developed as a tool  that allows to
vary and optimize all elements and the geometry (in 3-D)  of the beam
line providing  the results in terms of neutrino spectra and
distributions at large  distance with high statistics and in short
time. As a by-product, the  program provides also spectra and
distributions of secondary hadrons and muons  along the beam-line and
in the muon monitor pits after the hadron stop at the  end of the
decay tunnel.

The underlying idea is that in order to produce rapidly a neutrino
spectrum at large distance over a small solid angle (typically
$d\Omega \simeq 10^{-10}$  for the future LBL beams of CERN and FNAL),
one has to force all the mesons  to decay emitting a neutrino, and
force all neutrinos to cross the  detector volume. A weight is then
assigned to each neutrino,    proportional to the probability that
this process actually happened.

In practice this method is implemented by subdividing the simulation
into  four subsequent steps (as described in detail in the
following). The weight  assigned to each neutrino is the product of
factors originated in each  step times the solid angle subtended by
the detector:  
\begin{equation}
\nonumber
W_{tot} = \prod_i W_i  d\Omega_{det}.
\end{equation}

\subsubsection{Mesons production along target}

First the total number of secondary mesons ($\pi^+$, $\pi^-$, $K^+$,
$K^-$  and $K^0_L$) to be generated is calculated by integration of
the yield calculated in our model over the interesting range of 
longitudinal and transverse  momenta. These numbers obviously depend 
on target material, density and  length and on the number of protons 
on target.

Secondary mesons are generated along the target according to the
distribution of proton interaction points. The latter depends on
the  proton beam size and divergence, on the target  thickness,
$T_{targ}$, and on the proton interaction  length, $\lambda_p$, of the
target material. Momentum and angular distributions of the mesons are
sampled from the proposed parameterization. The weight associated 
with this step is 
\begin{equation}
\nonumber
W_1 = (1 - e^{-T_{targ}/\lambda_p}) e^{-z/\lambda_p}
\end{equation}
because protons are forced to interact in the target.

Meson trajectories in the target are calculated and their length,
$z'(z,\theta)$ used  to estimate the probability that the mesons 
exit the target without re-interacting: $e^{-z'/\lambda_s}$, 
where $\lambda_{s}$ is the meson interaction length in 
the target.

Tertiary contribution is added following the parameterization
described in Section \ref{yields}:
\begin{equation}
\nonumber
W_2 = e^{-z'/\lambda_s}
        \left[ 1+A_h(x) \int_0^z \frac{dy}{\lambda_p}w_z(y) \right]
\end{equation}
where $A_h(x)$, $w_z(y)$ and the approximation related to this
approach have been discussed above. In case of the segmented target considered
for the CNGS beam, this model gives the results of figure \ref{figyield}.

\subsubsection{Meson tracking in the neutrino beam-line}

The trajectory of each meson in the beam-line is calculated, taking
into account the tracking in the magnetic field of horn and
reflector, until it hits the walls of the  decay tunnel or the
collimators. The amount of material crossed by  the particle is also
recorded.

The meson is forced to decay along its trajectory, $traj$,
accordingly to its decay  length, $\lambda_{dec}$. The weight
associated to this process is  
\begin{equation}
\nonumber
W_3 = (1 - e^{-traj/\lambda_{dec}}) e^{-Z_{int}/\lambda_s}
\end{equation}
where $Z_{int}$ is the amount of material crossed up to the decay 
point and $\lambda_s$ is the interaction length in that material.

Contributions due to reinteractions of secondary particles 
in the material along the beam line are taken into account 
with a parameterization similar to that used for the target 
but in the ``short target'' approximation. 
The corresponding weight is:
\begin{equation}
\nonumber
W_4 = 1+A_h^{ter} (1-x)^{b_h^{ter}} \frac{Z_{int}}{2 \lambda_p} 
\end{equation}
as derived in formulae (\ref{linear}) and (\ref{terzeq}) of 
Section \ref{yields}

\subsubsection{Neutrino production from mesons}

For each meson a neutrino is produced; its flavour and its momentum
distribution in the parent meson rest frame depend on the decay  mode
and branching ratio, $B.R.$. The neutrino direction in the laboratory
frame is determined requiring that it  crosses the detector
volume. The angle, $\theta_{s\nu}$, between parent meson and neutrino
directions allows to calculate the Lorenz boost of the neutrino  from
the meson rest frame to the laboratory  frame. This in turns allows to
obtain the neutrino momentum in the  laboratory frame.

The weight associated to this process is proportional to the
probability  that the neutrino is emitted in the detector
direction. This is obtained  by simply boosting back the solid angle
subtended by the detector in the  meson rest frame, where the neutrino
is emitted isotropically;  
\begin{equation}
\nonumber
W_5 = B.R. \times (m_s / (E_s - P_s \cos\theta_{s\nu}))^2
\end{equation}
where $m_s$, $E_s$ and $P_s$ are the mass, energy and momentum of the 
secondary meson.

\subsubsection{Neutrino production from muons}

Muons are produced in the decay of secondary mesons taking into
account  the correct kinematics (branching ratio and
polarisation). Muons are also  tracked through the neutrino beam line
and forced to decay to produce   neutrinos in the detector
direction. An additional weight, $W_{3\mu}$ (equivalent
to $W_3$ for meson decays), is  introduced for the
neutrinos from muon decay. A weight $W_{5\mu}$ replaces $W_5$ 
accounting for the the muon decay kinematics (including polarisation).

\subsection{Statistical accuracy}

The statistical accuracy of this way of simulating neutrino beams does
not depend much on the  distance between the detector position and the
neutrino source as it is the  case for classical unweighted methods.

In the classical unweighted case, only a fraction of the pions
(typically 5--10\%) decay before interacting (either in the
beam-line material or in the decay tunnel walls); in addition 
neutrinos are spread over a wide solid angle (about 1 mrad in the
CNGS case) because of the decay kinematics. It follows that, 
to enhance statistics, at large distance the neutrino spectra need to 
be computed on a surface
much wider than  the actual detector area, relying on the fact that
the spectra shapes  varies slowly with the radius. In the CNGS case
an accuracy of a few percent can be achieved with several millions 
protons on target if a detector area larger than  $\simeq 10^4$~m$^2$ is 
considered.

In the parameterization case, since all mesons -- within the focusing
optics acceptance -- are exploited to produce  neutrinos in the detector,
the statistical accuracy is independent from  the detector distance
and proportional to the inverse of the square  root of the number of
generated positive pions (for $\nu_\mu$ beams),  namely about the
number of generated proton interactions on target. An accuracy  better
than a percent is thus obtained with less than $10^5$ p.o.t., for any
size of the detector surface.
 
\subsection{Simulations of past and future neutrino beams}

In order to give a more quantitative appreciation of the accuracy that
one can obtain in the simulation of neutrino beams using the generator
described in the previous Sections, we present the comparison with the
published neutrino spectra measured with the CHARM II detector 
\cite{CHdet} exposed at the CERN-WANF beam.

In addition we present the comparison between the CNGS simulation
based on our method and that based on the FLUKA stand-alone program for 
secondary meson production interfaced with GEANT for particle tracking.

\subsubsection{Comparison with CHARM II data}

The WANF neutrino beam line at CERN, to which the CHARM II detector 
was exposed, is well described elsewhere \cite{WANF}. Its main 
characteristics has been summarized in Table~\ref{tab:ngstab}.
The facility was run, during several years of operation, either 
selecting  positive charged particles ($\nu_\mu$ beam) or negative 
ones ($\overline \nu_\mu$ beam). Neutrino/anti-neutrino
interactions were collected in the CHARM II detector and fully
reconstructed \cite{CHARM}.

\begin{figure*}
\begin{center}
\mbox{\epsfig{file=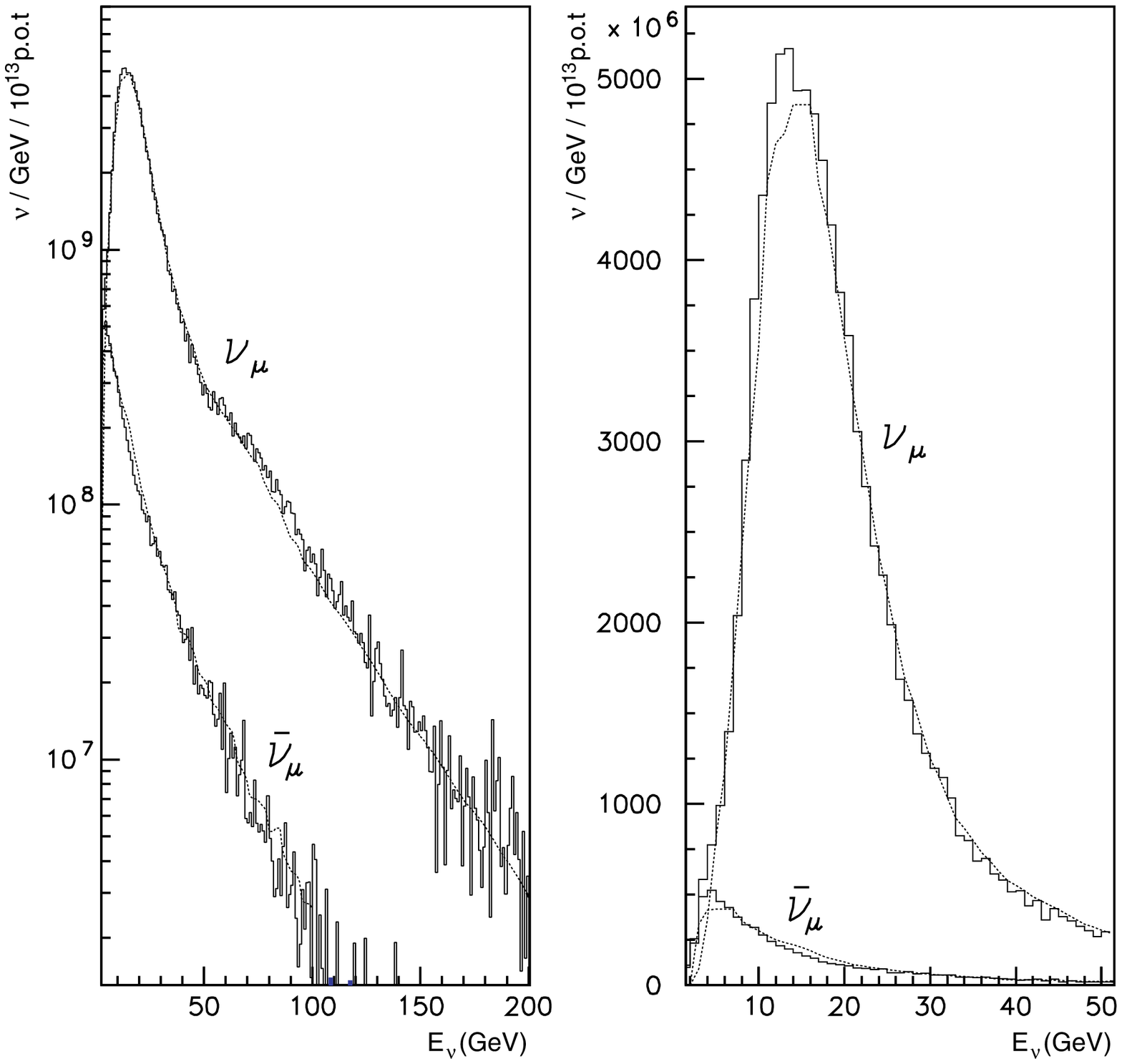,width=0.75\linewidth,height=0.45\textheight}}
\mbox{\epsfig{file=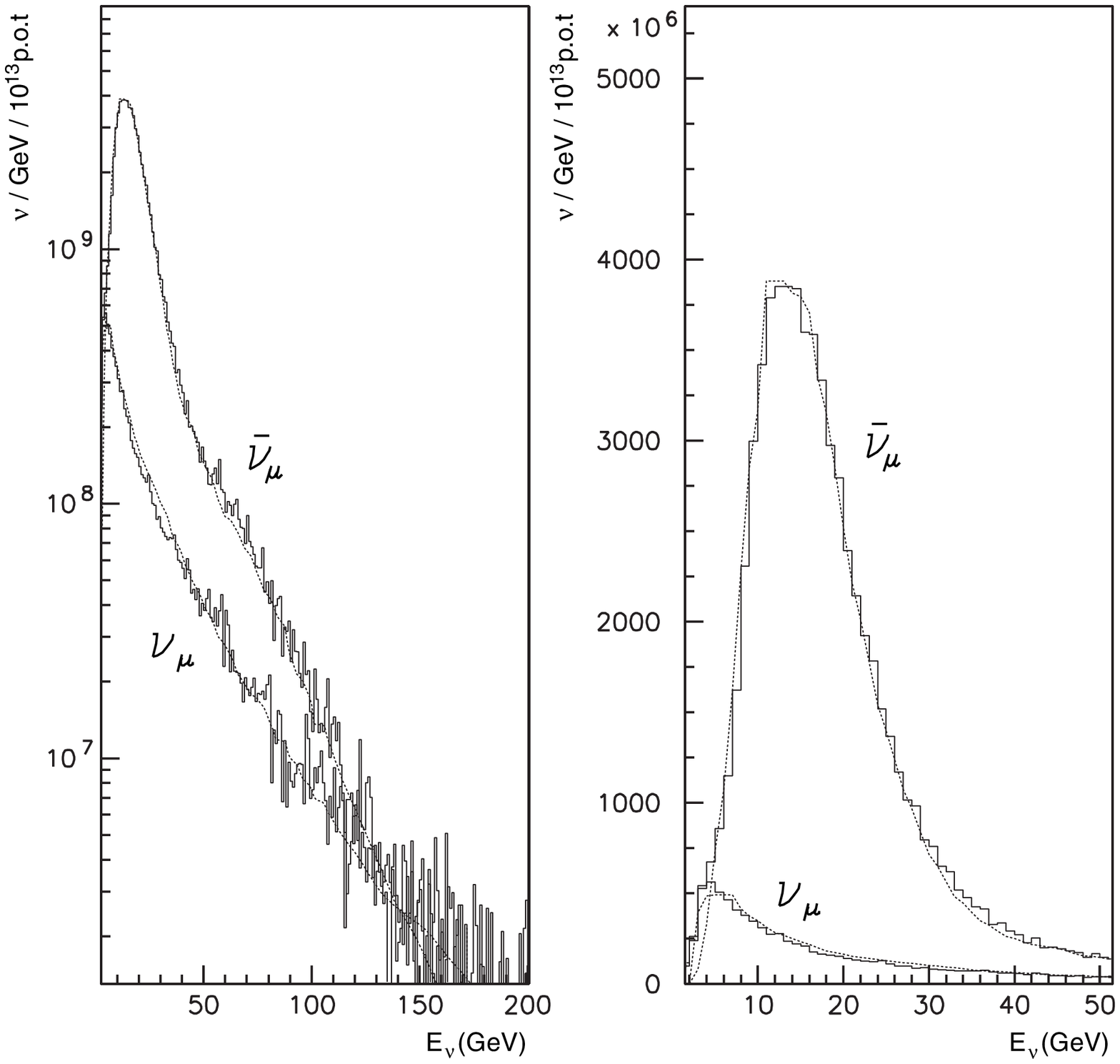,width=0.75\linewidth,height=0.45\textheight}}
\end{center}
\caption{The WANF neutrino ({\it top}) and anti-neutrino ({\it bottom}) 
fluxes at the CHARM II detector: the dotted lines are experimental data 
from Ref. \cite{CHARM}, the continuous line is the beam simulation. 
On the left, logarithmic scale is used to make evident the spectral 
behaviour at high energy as well the wrong sign contamination; on the 
right, linear scale is used for a better appreciation of the 
focusing/defocusing effect.}
\label{fig:nu_char}
\end{figure*}

In figure Figure~\ref{fig:nu_char}--{\it top}, we show the comparison 
between the measured neutrino fluxes in case of positive mesons focusing, 
($\nu_\mu$ beam with $\overline \nu_\mu$ contamination) and the 
simulation performed with our method and based on
$10^5$ p.o.t.. On the left hand side a logarithmic scale is used to 
make evident the spectral behaviour at high energy as well as the wrong
sign contamination; on the right hand side a linear scale is used for
a better appreciation of the focusing/defocusing effect.
In figure  Figure~\ref{fig:nu_char}--{\it bottom}, the same plots are shown 
in the case of negative meson focusing ($\overline \nu_\mu$ beam with 
$\nu_\mu$ contaminations). 

The overall agreement is at the percent level, with at most 10\%
disagreement on a bin per bin basis.

The high energy tails of the distributions are dominated by the
production of high energy secondary mesons peaked in the forward
direction, and are  practically  insensitive to the magnetic
focusing. In fact high energy mesons, with small angular aperture,
travel most likely through the neck of the horn where they are hardly
deviated. The good agreement between simulation and data indicates
that high $x_F$ production on target is well simulated and that
re-interactions on the material along the beam-line is correctly 
taken into account.

In the focusing/defocusing energy range the agreement is an indication
that low $x_F$ production is correctly generated at least up to
$\simeq 10$ mrad (the WANF optics acceptance). The fact  that also the
wrong sign contamination in the simulation behaves as the data, means
that tertiary production in target and down-stream material (mainly
the horn neck) is described to a sufficient level of approximation. 

\subsubsection{The CNGS beam-line}

\begin{figure*}
\begin{center}
\mbox{\epsfig{file=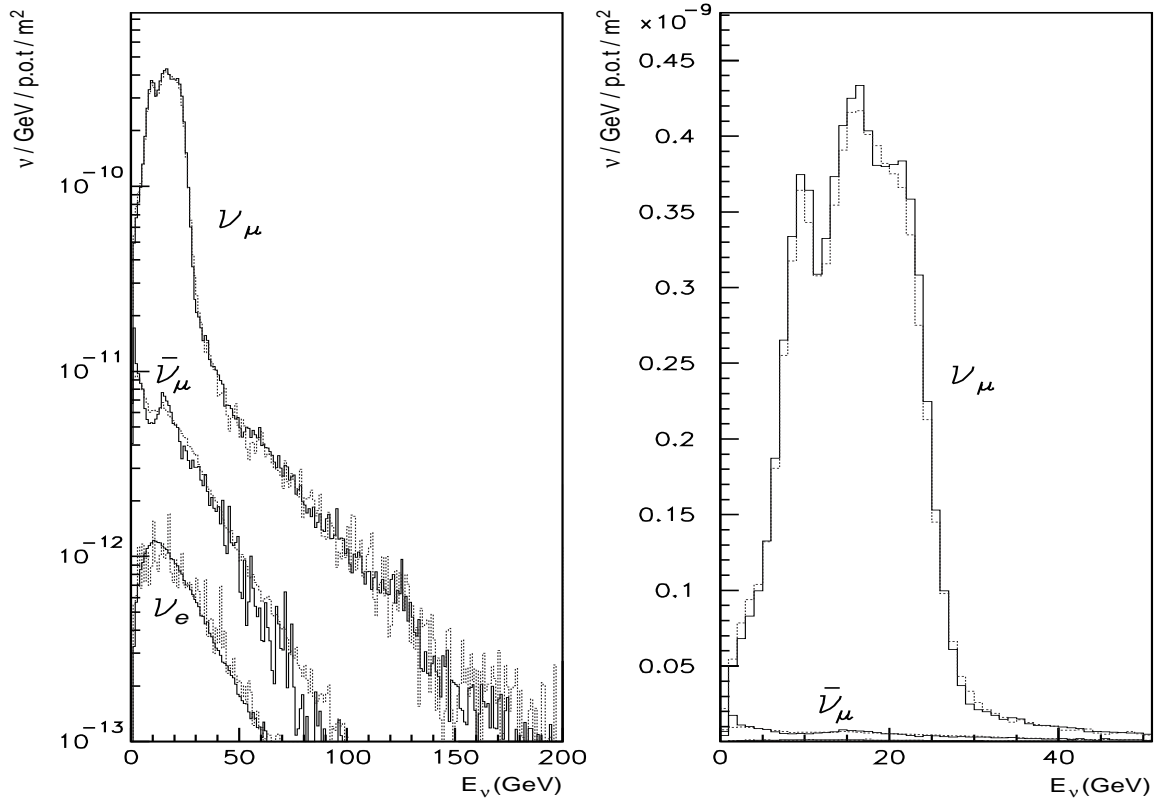,width=0.85\linewidth,height=0.45\textheight}}
\end{center}
\caption{The CNGS neutrino fluxes at Gran Sasso (732 Km from target).
Comparison between FLUKA/GEANT simulation (the dotted lines) and the 
parametrised simulation (continuous line)
On the left, logarithmic scale is 
used to make evident the spectral behaviour at high energy 
 as well the wrong sign
contamination; on the right, linear scale is used for a better  
appreciation of the focusing/defocusing effect.}
\label{fig:nu_cngs}
\end{figure*}

As mentioned earlier, our neutrino beam generator was originally
developed  to allow a rapid optimization of LBL neutrino beams. To test
the reliability of it, we have performed a detailed simulation of the
CNGS LBL  neutrino beam for an extensive comparison with the full beam
simulation based  on the FLUKA stand-alone package \cite{FONE} for 
secondary particle production and on GEANT for tracking along the beam 
line \cite{CNGS}.

Apart from the tunnel geometry and the focusing optics, the main
differences of the CNGS beam-line with respect to the WANF are
the target material ($carbon$ instead of $beryllium$) and the
proton beam  energy ($400$ GeV instead of $450$ GeV). These differences
have been  taken into account in our simulation with the scaling laws
proposed and described in Section 4.2 and 4.3.

As remarked in Section 4.3, the agreement at the target level between 
our calculation and the FLUKA based full Monte Carlo program 
is noticeable, provided that tertiary production is taken into 
account in the former case. The slight disagreement visible in the 
$p_T$ distributions of all charged particles mainly at large $p_T$ 
translates in a slightly more forward peaked distribution of the exit 
points along the target.

Remarkably enough, as far as the neutrino spectra at large distance
are concerned, the discrepancies between the two models do not
propagate with the same strength (see Figure~\ref{fig:nu_cngs}). 
This is because in the momentum range and angular acceptance selected 
by the focusing optics both particle production models reproduce 
very well the available experimental data. 

Before concluding, it is worth mentioning that early simulations of 
the CNGS beam line based on GEANT stand-alone disagreed with those 
presented here, being too optimistic by more than $\simeq 20\%$. 
On the other hand FLUKA stand-alone gives results fully compatible 
with those presented here. 

\section{Conclusions}

Empirical formulae for single-particle inclusive invariant cross 
sections in $p$-Be interactions have been derived, on the basis 
of single-particle inclusive production data collected by the 
NA20 \cite{NA20} and NA56/SPY \cite{SPY1,SPY2} experiments. 
These formulae reproduce the experimental data within a 10\% 
accuracy. 

The hypothesis of Feynman scaling has been verified to hold 
with our parametrisation giving a suitable representation of 
production data collected over a wide range of primary proton beam 
energies (from 24 GeV to 450 GeV). 

Prescriptions to extrapolate this parameterization to finite 
targets and to targets of different materials have been given. 

The results obtained have been used as an input for the simulation 
of neutrino beams. A comparison to data collected by the CHARM-II 
neutrino experiment at CERN has demonstrated the capability of 
this approach to predict the past. 

These formulae can be of great practical importance for fast 
calculations of neutrino fluxes and for designing new neutrino 
beam-lines. Thus they may be used in fast simulations aiming at 
the optimisation of the long-baseline neutrino beams at CERN and 
FNAL. Predictions for the neutrino spectra of the CNGS beam from 
CERN to Gran Sasso have also been given.

\section*{Acknowledgements}
We gratefully acknowledge enlightening discussions with and 
fruitful suggestions of K. Elsener, L. Gatignon, P.G. Ratcliffe, 
S.I. Striganov. We are indebted to P.~Gorbunov, for having 
provided us the files with neutrino spectra measured in 
CHARM II. We are grateful to S. Ragazzi, our spokesman in the 
NA56/SPY collaboration, and P.~Picchi for their encouragement 
to this work and to all our colleagues in the NA56/SPY 
collaboration.

\end{document}